\documentclass{aa}

\usepackage{graphicx}
\usepackage{txfonts}
\usepackage{xcolor}

\usepackage{float}

\usepackage{natbib}
\usepackage{hyperref}
\usepackage{graphicx,dblfloatfix}
\definecolor{purple}{rgb}{0.5,0,0.87}

\hypersetup{colorlinks,linkcolor={black},citecolor={blue},urlcolor={purple}} 
\usepackage{adjustbox}
\usepackage{booktabs}
\usepackage{float}
\usepackage{subcaption}
\usepackage{placeins}

\hypersetup{colorlinks,linkcolor={cyan},citecolor={blue},urlcolor={purple}} 

\newcommand{\Msun}{M$_{\odot}$}

\makeatletter
\renewcommand{\p@subfigure}{}
\makeatother

\renewcommand\thesubfigure{\thefigure\alph{subfigure}}

\begin{document}

   \title{Bridging integrated-light and resolved-star studies\\of Local Group dwarf spheroidals}
   \subtitle{A novel use of deep imaging based on amateur telescopes}
   
   \author{Sergio Guerra Arencibia \inst{1,2}\corrauth{sguerra@iac.es} 
           \and
           Ignacio Trujillo \inst{1,2}
           \and
           Mireia Montes \inst{3}
           \and
           Ignacio Ruiz Cejudo \inst{1,2}
           \and
           Carlos Marrero-de la Rosa \inst{1,2}
           \and
           Vicente Fontana
           \and
           Aleix Roig \inst{4}
          }

    \institute{Instituto de Astrof\'isica de Canarias, c/ V\'ia L\'actea s/n, E-38205 - La Laguna, Tenerife, Spain  \email{}
    \and
    Departamento de Astrof\'isica, Universidad de La Laguna, E-38206 - La Laguna, Tenerife, Spain
    \and
    Institute of Space Sciences (ICE, CSIC), Campus UAB, Carrer de Can Magrans, s/n, 08193 Barcelona, Spain
    \and
    Parc Astronòmic Muntanyes de Prades, C/ Muralla, 3 43364, Prades, Tarragona, Spain
    }

   \date{}

  \abstract
   {Low-mass galaxies beyond the Local Group can only be studied through their integrated light, yet it remains unclear whether structural parameters inferred from integrated light are directly comparable to those obtained from resolved-star studies. Using new deep imaging obtained with small-aperture amateur telescopes, we analyse the classical dwarf spheroidals Leo II, Sculptor, and Fornax, whose stellar distributions have been extensively characterised through resolved-star studies. We find that the radial light distributions and effective radii inferred from integrated light are in excellent agreement with those derived from resolved stars, showing that resolved-star and integrated-light analyses can be placed within a common observational framework applicable across a broad range of distances. The two approaches are highly complementary: resolved-star analyses trace galaxies to very large radii but are limited by stellar crowding in their central regions, whereas integrated-light observations are insensitive to crowding and allow the inner stellar distribution to be traced to substantially smaller radii. The integrated-light analysis extends the structural analysis to radii up to an order of magnitude smaller than previous star-counting studies while significantly reducing the scatter in the measured profiles.}


   \maketitle

\section{Introduction} \label{Sec:intro}

Dwarf galaxies are key laboratories for understanding galaxy formation and evolution at the lowest-mass scales, where the $\Lambda$ cold dark matter ($\Lambda$CDM) cosmological model faces several challenges \citep{bullock2017}. Within the Local Group, and especially surrounding our Galaxy, these low-mass objects have long served as benchmarks for studies of, amongst others, galaxy formation mechanisms, star formation histories, chemical evolution, and dark matter physics \citep[e.g.][]{Mateo1998, McConnachie2012}. The advantage of these systems is that they are so close to us that their stellar bodies are resolved into stars \citep[e.g.][]{Mateo1998, Tolstoy2009}, enabling analyses based on photometry, spectroscopy, and kinematics of individual stars. On the other hand, modern very deep surveys such as LIGHTS \citep{Trujillo_lights2021, zaritsky2024}, LSST \citep{Ivezi2019, rubidr1}, or Euclid \citep{euclid} are already identifying, and will continue to identify, large numbers of low-mass systems, some of them analogues to the local objects in luminosity and mass. However, at their distances and with current imaging resolution, we can only study their integrated light.

As a result, two approaches have emerged: analyses based on individual stars, which are only applicable to nearby resolved systems, and integrated-light analyses, the only option for distant unresolved ones. In practice, however, the transition between these regimes is not abrupt. At intermediate distances, systems fall within  the semi-resolved regime \citep[e.g.][]{renzini1998,conroy2016,fastar1}, where stellar populations are only partially resolved, and both approaches are in principle applicable---though the application of integrated-light techniques in this regime remains poorly understood.

\begin{figure}[htb]
    \centering
    \includegraphics[width=0.9\linewidth]{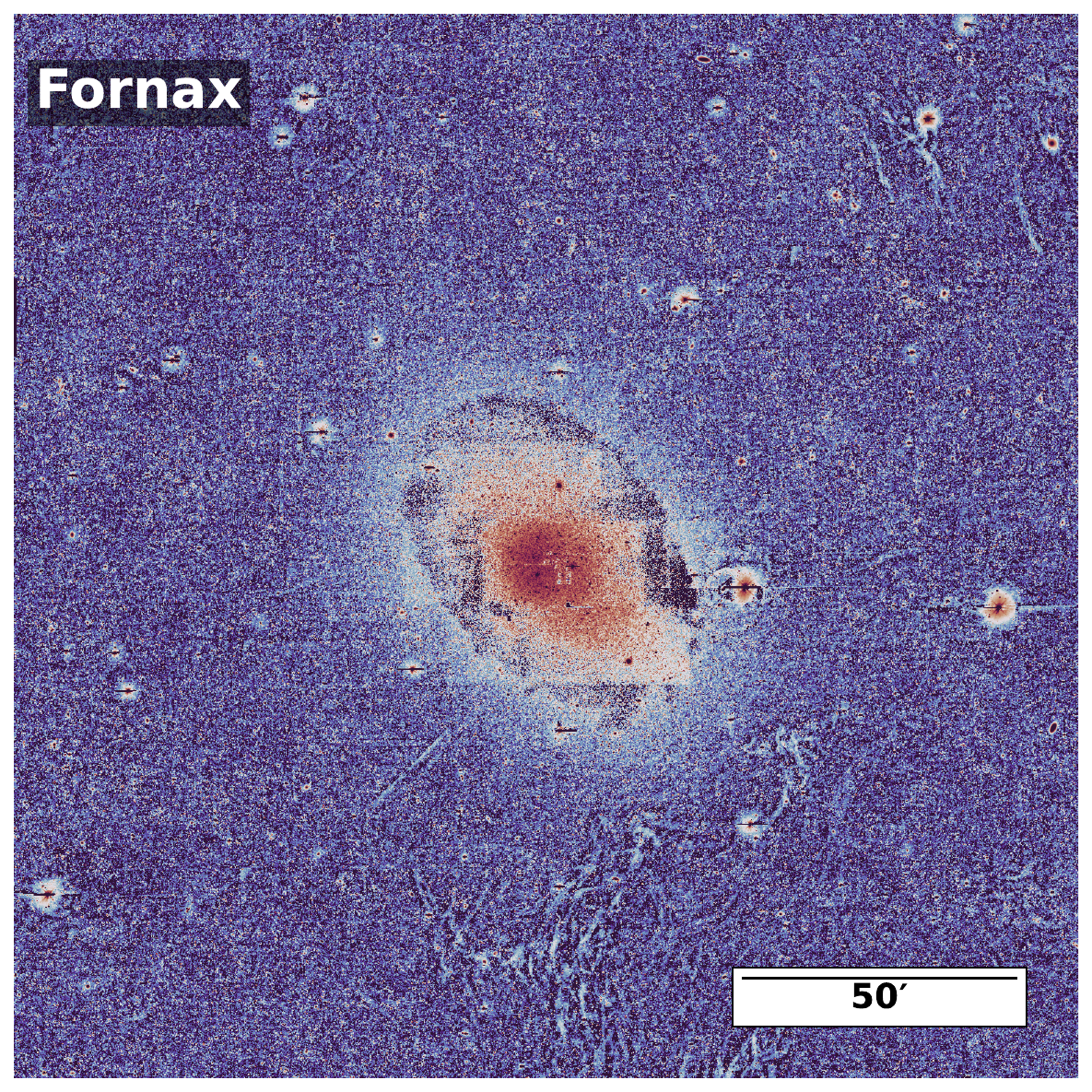}
    \caption{Fornax dwarf spheroidal in the \textit{g} band from Legacy Surveys DR10 (DES; \citealt{DECaLS}). The  visualisation parameters have been tuned to highlight the low-surface-brightness structure. The diffuse stellar emission from the galaxy and the Galactic cirrus emission surrounding it have not been preserved in the reduction process, resulting in the loss of information and artefacts.}
    \label{fig:FornaxDecals}
\end{figure}

Bridging these two approaches and comparing the results yielded by each is a necessary step for advancing our understanding and safely combining all the available information of low-mass objects. A natural approach would be to use existing observations in the semi-resolved regime, which are already being used for resolved-star studies, and to perform integrated-light analyses on them. Thus, we could assess the agreement between the inferred structural parameters using the two techniques. One option to carry out this study would be to use modern deep surveys; Fig.~\ref{fig:FornaxDecals} shows the Fornax dwarf spheroidal in one of them (the \textit{g} band from Legacy Surveys DR10, specifically from the Dark Energy Survey; \citealt{DECaLS}) with the visualisation parameters adjusted to highlight the low-surface-brightness features. The diffuse stellar emission has been over-subtracted during the reduction process, effectively removing information from the data and introducing artefacts \citep[e.g.][]{TrujilloAndFliri2016, Kelvin2012}. Therefore, performing an integrated-light analysis using these data from modern deep surveys with professional telescopes is not straightforward.

This over-subtraction problem plagues modern deep surveys and stems from the fact that the field of view (FOV) of the cameras is covered by many CCD detectors. These detectors are smaller than the angular size of the target galaxy, and when this occurs, the detector is fully covered by signal from the galaxy, preventing an accurate sky estimation. As an example, Fornax has an effective radius ($r_{\rm{e}}$) of $\sim 16$ arcmin \citep{McConnachie2012}, whereas each DECam \citep{decam} detector covers $18\times9$ arcmin and each LSSTCam \citep{lsstcam} detector covers $13.3\times13.3$ arcmin. One possible workaround, and the one adopted in this work, is to use telescopes with a large FOV but large monolithic detectors, such that the detector always extends beyond the galaxy, leaving pixels free of signal for an accurate characterisation of the sky.

We aim to establish the bridge between studies based on individual stars and integrated light. We obtain new deep imaging of three classical dwarf spheroidal galaxies (Leo~II, Sculptor, and Fornax) using small-aperture amateur telescopes and analyse them on an integrated-light basis. Then, by directly comparing the radial distribution of light obtained with both approaches, we assess the agreement of the measurements inferred from both techniques. This comparison provides a stepping stone for placing observations of low-mass galaxies across a broad range of distances into a unified framework. We also take advantage of the fact that our targets fall in the semi-resolved regime, allowing us to characterise the challenges that integrated-light analyses face when applied in this regime.

The paper is structured as follows. Sec.~\ref{sec:Data} provides an overview of the data, observations, and data reduction. Sec.~\ref{Sec:analysis} describes the analysis and presents the results, which are then discussed in Sec.~\ref{sec:Discussion}. Sec.~\ref{sec:Conclusions} provides the conclusions of the study. Throughout this work, magnitudes are given in the AB system unless otherwise specified and the radial profiles are given along the semimajor axis.

\section{Data}\label{sec:Data}

\begin{figure*}[h]
    \sidecaption
    \centering
    \includegraphics[height=5.4cm, keepaspectratio]{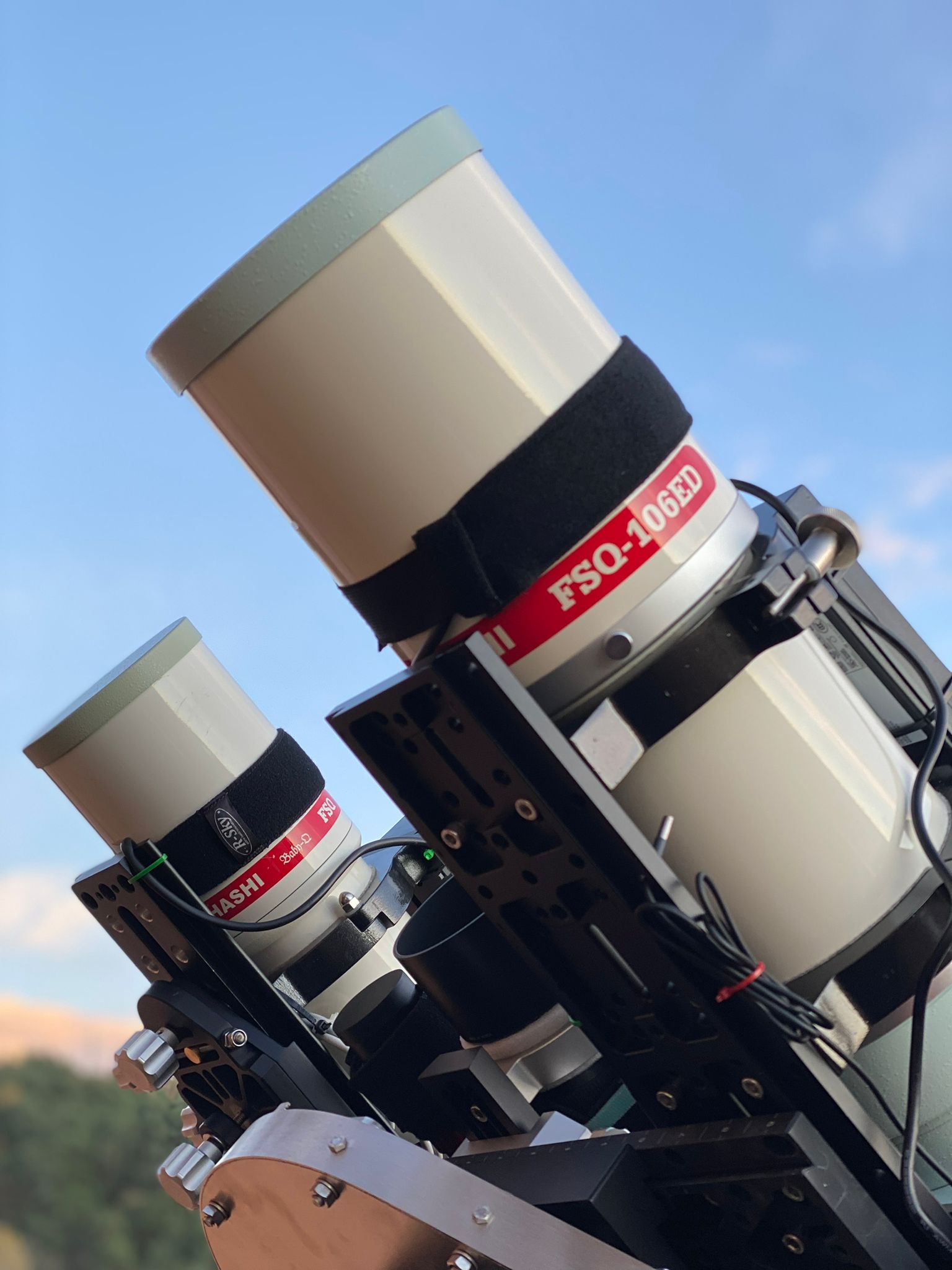}
    \hspace{0.5cm}
    \includegraphics[height=5.4cm, keepaspectratio]{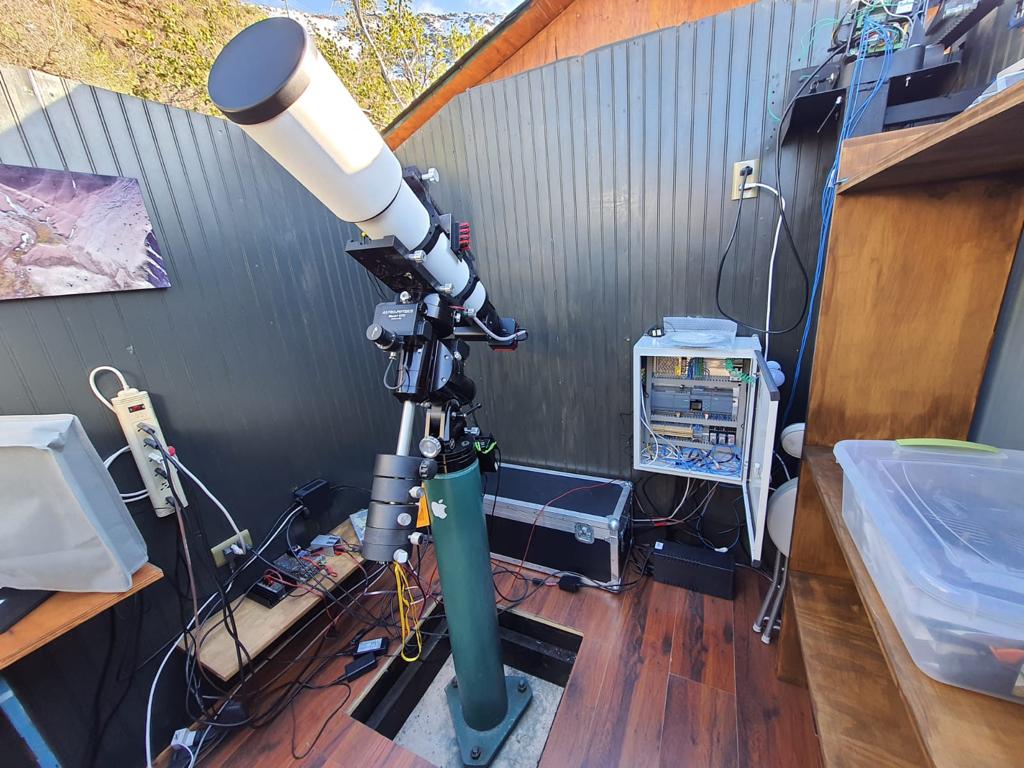}
    \caption{Telescopes used to perform the observations presented in this work. Left: the Astrocat telescope ($D\sim10.6\ \rm{cm}$), which observed the Leo~II dwarf spheroidal. Right: the AstroRemoto telescope ($D\sim13\ \rm{cm}$), which observed the Sculptor and Fornax dwarf spheroidals.}
    \label{fig:telescopesImages}
\end{figure*}

The data used in this work have been obtained using the small-aperture ($\rm{D}\sim10\ \rm{cm}$) amateur telescopes shown in Fig.~\ref{fig:telescopesImages}. The observations of Leo~II were performed between February and March 2024, while the observations of Sculptor and Fornax were performed between September and December 2025. The data were obtained using a Luminance filter (the same in both telescopes), which covers the wavelength range between 4000 and 7050 \r{A}. This luminance filter is wider than the typical V-band but centred at the same wavelength; consequently, for the sake of clarity, we have computed the transformation factor between filters so that the magnitudes given throughout this work are in the Bessell V-band (details in App. \ref{App:transmittances}).  For the colour images, observations with the filters R, G, and B were performed for Sculptor and Fornax, integrating one hour in each filter. However, the colour image of Leo~II is built using the Sloan Digital Sky Survey DR16 (SDSS, \citealt{sdssdr16}) data. This is because Leo~II was observed prior to the adoption of multi-band imaging in this campaign, and the availability of SDSS made dedicated R, G, and B exposures unnecessary. The transmittance curves of the filters are shown in App. \ref{App:transmittances}. The total exposure time, full-width-at-half-maximum (FWHM), and surface brightness limits of the luminance stacked images are reported in Table \ref{tab:depths}. The surface brightness limits are provided as $3\sigma$ fluctuations of the background of the images in areas equivalent to $1^{\prime}\times1^{\prime}$. We use this metric instead of the traditional $10^{\prime\prime}\times10^{\prime\prime}$ \citep[see e.g.][]{TrujilloAndFliri2016, roman2020} to better suit the enormous extensions in the sky ($\sim1\ \rm{deg}$) of the objects under analysis.

\begin{table}[h]
\centering
\caption{Summary of the final stacked images.}
\begin{adjustbox}{width=0.8\linewidth}
\begin{tabular}{c cccc}
\toprule
Target & $t_{\rm{exposure}}$ & FWHM & $\mu_{\rm{limit}}$ ($3\sigma, 1^{\prime}\times1^{\prime})$ \\
 & (h) & (arcsec) & ($\rm{mag}/\rm{arcsec}^2$) &    \\
\midrule
Leo~II    &  58.0 & 3.1 & 30.1 \\
Sculptor &  28.5 & 2.8 & 29.8 \\ 
Fornax   &  26.0 & 2.5 & 29.6 \\
\bottomrule
\end{tabular}
\end{adjustbox}
\label{tab:depths}
\end{table}

\subsection{Astrocat observatory telescope}

Astrocat observatory obtained the data of Leo~II dwarf spheroidal. The telescope is located approximately at $41^{\circ}18'42.0''N\,\,\,0^{\circ}59'24.0''E$ (around $40\,\rm{km}$ to the north-west of Tarragona, Spain) and at an altitude of $\sim975\,\rm{m}$ above sea level. The telescope has an aperture of 106 mm and a focal length of $530\,\rm{mm}$ at f/5. The camera used was the \texttt{ASI 2600 MM Pro Mono}, which has $6248\times4176$ px, resulting in a FOV of $\sim2.5 \times 1.7\,\rm{deg}^2$ at a pixel scale of $1.464\,\rm{arcsec}/\rm{px}$.

\subsection{AstroRemoto observatory telescope}

AstroRemoto observatory obtained the data of Sculptor and Fornax dwarf spheroidals. It is approximately located at $32^{\circ}42'59.8''S\,\,\,70^{\circ}28'59.9''W$ (around $100\,\rm{km}$ to the north of Santiago de Chile) and at an altitude of $\sim1650\,\rm{m}$ above sea level. The telescope has an aperture of 130 mm and a focal length of $655\,\rm{mm}$ at f/5. The camera used was the \texttt{ASI 1600 GT Mono}, which has $4656\times3520$ px, resulting in a FOV of $\sim1.5 \times 1.1\,\rm{deg}^2$ with a pixel scale of $1.164\, \rm{arcsec}/\rm{px}$.

\subsection{Observational strategy}

Deep imaging demands a careful characterisation of the illumination of the sky at the moment of performing the observations. This requires a specific observational strategy, since twilight and dome flats do not account for the illumination of the scientific data and introduce their own gradients. Thus, we use the science exposures themselves to build the flat-field, this way correcting at once the detector sensitivity and the illumination of the observed field. To achieve this, we followed an observational strategy similar to that described in \cite{TrujilloAndFliri2016}, where the dithering pattern step is of the order of the extension of the galaxy. These large dithering steps allow us to always get pixels free of signal in at least one exposure, characterising this way the detector sensitivity and the illumination of the field. 

The exposure time of the individual frames is a compromise between being large enough to get enough signal in each exposure and minimise overheads, and small enough to characterise the illumination of the sky via the dithering pattern in a reasonable time window. Given the properties of the telescopes and the angular size of the objects, the exposure time has been set to 180 seconds.

\subsection{Data reduction}

A low-surface-brightness compliant pipeline has been used to reduce the data, aiming to preserve the low-surface brightness features as good as possible. Similar data reductions can be checked in \cite{Trujillo_lights2021, Golini2024, junais2025}. The following subsections provide a description of the steps performed in the reduction process.

\subsubsection{Bias and dark current correction}

First, in order to remove pixels that are not in the dynamic range of the detector, we mask every pixel with a value greater than $65500\, \rm{ADUs}$. Then, the raw frames need to be corrected from bias and dark current. For each night a master-dark (which also contains the bias) is built by combining the individual darks. Then these master-darks are subtracted from the individual frames.

\subsubsection{Flat-field and illumination correction}

Once the data are corrected from bias and dark current, we correct for the different sensitivity of the pixels across the detector and for the illumination of the observed field. As noted by  \cite{TrujilloAndFliri2016}, building the flat-field with the scientific data itself (i.e. using the background regions of the scientific data as the illumination) allows us not only to avoid inhomogeneities and gradients present in twilight and dome flats but also to correct the illumination of the field.

For building the flat-field with this methodology, we first need to normalise all the frames to the unit. To perform this normalisation properly, the illumination pattern of the camera has to be taken into account. We use an annulus located at the region that is expected to be illuminated with a common efficiency, avoiding the central region where the target is typically placed and the outermost regions affected by vignetting. Computing the median value within this region allows us to normalise the individual frames. Next, the normalised images are combined with a sigma clipping, obtaining a preliminary flat-field. The improvement on the images after applying this preliminary flat-field is significant, and a large amount of previously hidden signal is now detectable. Identifying this hidden signal allows us to build masks that are applied to the bias-dark corrected images in order to repeat the process and build a new flat-field that includes less contamination. This process is repeated two more times to maximise the amount of identified signal, obtaining a final flat-field. Additionally, not all the images are used together to build this flat-field. Sky illumination varies with time, so to characterise it more accurately, we build the flat-field using frames close in time to the frame that is being corrected \citep[as shown in][ for example]{Saremi2025}. The number of frames used for this ‘running flat’ strategy depends mainly on the number of background counts collected in each exposure. We found 20 (i.e. using the previous and following ten frames for building each flat-field) to be optimal. Finally, to minimise vignetting towards the edges of the images, the pixels whose values in the flat-field are below 0.85 or above 3 were masked.

\subsubsection{Astrometry, sky subtraction, and photometric calibration}

The next step in the reduction of the data is to assign to each pixel its position on the sky (i.e. compute an astrometric solution). An initial astrometric solution is obtained using the utility \texttt{solve-field} from the package \texttt{astrometry.net} \citep{Barron2008}. This solution is later improved by employing \texttt{SExtractor} \citep{BertinArnouts1996} and \texttt{SCAMP} \citep{Bertin2006}, refining the astrometry and correcting for distortions. 

The estimation and subtraction of the background allow us to remove the contribution of the background signal from our images. This is a critical step for preserving the low-surface-brightness signal, and we choose to keep the characterisation of the sky as simple as possible, simply subtracting a constant from each image. To estimate this value, we first build an aggressive mask using \texttt{Noisechisel} \citep{Akhlaghi2015} and then compute the median value of the unmasked pixels.  

To convert from ADUs to physical units, we calibrate our images photometrically. To do so, we identify the stars in our images with $14 < \rm{m_{\rm{V}}} < 16$ which simultaneously have GAIA spectrum \citep{gaia2023}. We compute the magnitudes of the selected stars based on their spectra (by convolving them with the transmittance of the filter used to take the data), and we also measure their fluxes in our images (still in ADUs). After exploring different apertures for measuring the flux of the stars in our images, we conclude that using large apertures ($r=7r_{\rm{e}}$) is optimal, as they capture the full flux from each source without missing the wings of the point spread function (PSF). Once we have both measurements, we compute the factor that we need to apply to our images in order to get them calibrated. Since all the frames have been taken with the same instrumentation and configuration, we use the distribution of all the individual calibration factors (one per exposure) to compute a common factor, effectively reducing the noise in the calibration. The physical units of the images are set to nanomaggies, corresponding to a photometric zero-point of 22.5.

\subsubsection{Coaddition}

The methodology used for stacking the corrected frames is based upon the fact that not all the frames have the same quality, with lots of factors contributing to this (e.g. night conditions, air mass, light pollution). In this situation, combining all the frames equally will not result in the best stacked image possible, and a weighting scheme needs to be introduced in order to perform the combination optimally. Therefore, we adopt the standard deviation of the background as an indicator of the quality of the frame. Thus, the weight of the $i^{th}$ frame is the quadratic ratio of the standard deviation of the sky in the best exposure ($\sigma_{\rm{min}}^{2}$) to the standard deviation of the sky in the $i^{th}$ frame (i.e. $\sigma_{\rm{min}}^{2} /\sigma_{i}^{2}$). This weighting scheme is optimal for images with Gaussian noise \citep[see ][Chap. 4.4.5]{leo1988techniques}. During the stacking process, the pixels with undesired signal (e.g. cosmic rays and satellites streaks) should not be included. Therefore, a sigma-clipping rejection per pixel is applied to all the exposures that are going to be combined.

\begin{figure*}[h]
    \centering
    \includegraphics[width=\linewidth]{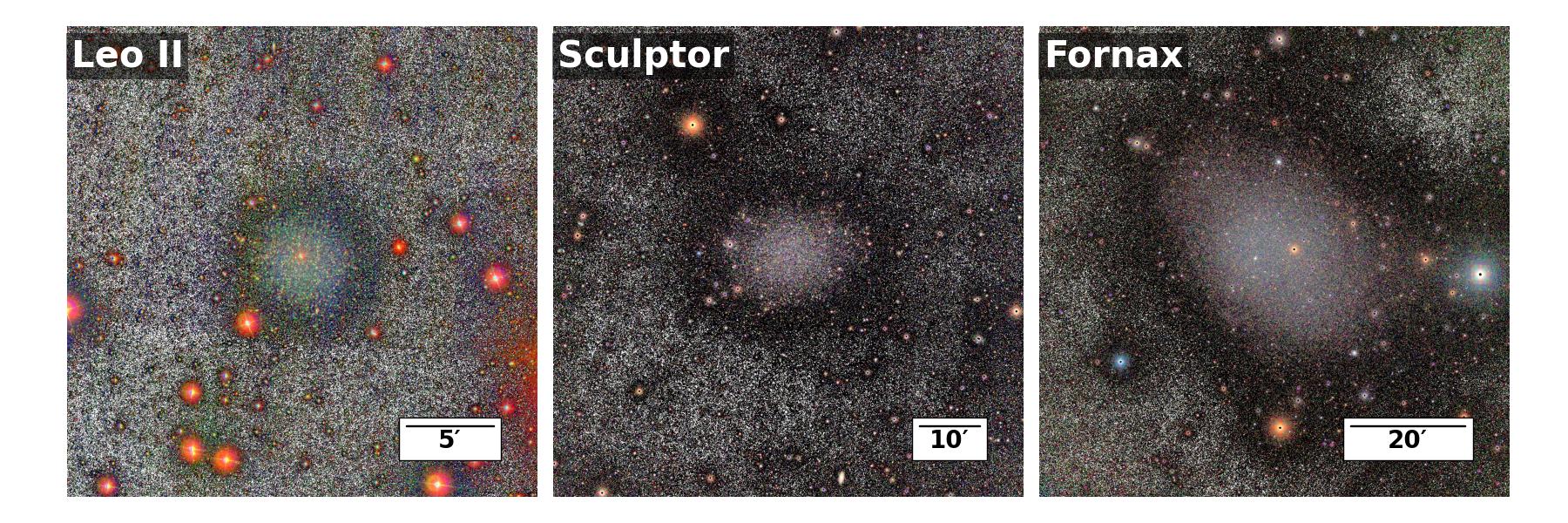}
    \caption{From left to right: colour images built using \textit{astscript-color-faint-gray} \citep{sainz2024} of Leo~II, Sculptor, and Fornax dwarf spheroidals. The colour regions of Leo~II use SDSS data, and the ones of Sculptor and Fornax use the observations presented in this work. The deep luminance data is shown on an inverted greyscale to maximise the visualisation of faint surface brightness features.} 
    \label{fig:colourImages}
\end{figure*}

At this point we have a coadded image which is deeper than any individual exposure, where lots of low-surface-brightness features and sources have now emerged from the noise. We take advantage of this and use the mask of the coadded image to improve the sky estimation and subsequent steps. This second iteration using the mask of the first coadded image produces the final data that is used for the analysis. The final images of the reduction process are shown in Fig.~\ref{fig:colourImages}. The figure shows a combination of the RGB images (Sec.~\ref{sec:Data}) and the deep luminance images on an inverted greyscale as background.

\subsection{Star subtraction}

Subtracting the scattered light from bright stars from the images is frequently a needed step to mitigate the contamination of undesired light \citep{sandin2014,TrujilloAndFliri2016}. In this work, the subtraction was performed on Fornax's and Sculptor's fields, where the scattered light plays a relevant role due to the large angular size of the galaxies and their proximity to bright stars; given the smaller size of Leo~II plus the presence of fewer bright stars in this field, we do not consider star subtraction necessary for this object.

We have constructed the extended PSF of our images following the prescriptions from \cite{marreroDeLaRosa2026, marreroDeLaRosa2026b}. The code LISAN\footnote{https://github.com/CarlosMDLR/LISAN} (Layered Intensity Spread and Analysis for Night-sky structures) was used to model the PSF using non-saturated stars, reaching up to 30\arcsec with high signal-to-noise ratio. From there, the PSF was extended using a clean region of the brightest star in the Fornax's image ($\rm{\lambda}^2$ For, located at RA: 02$^h$36$^m$58.60$^s$ and Dec: -34$^d$34$^m$44.86$^s$), exhibiting an outer power law in surface brightness with a slope of $\alpha = -2.03$. 

\begin{figure}[htbp]
    \centering
    \includegraphics[width=0.9\columnwidth]{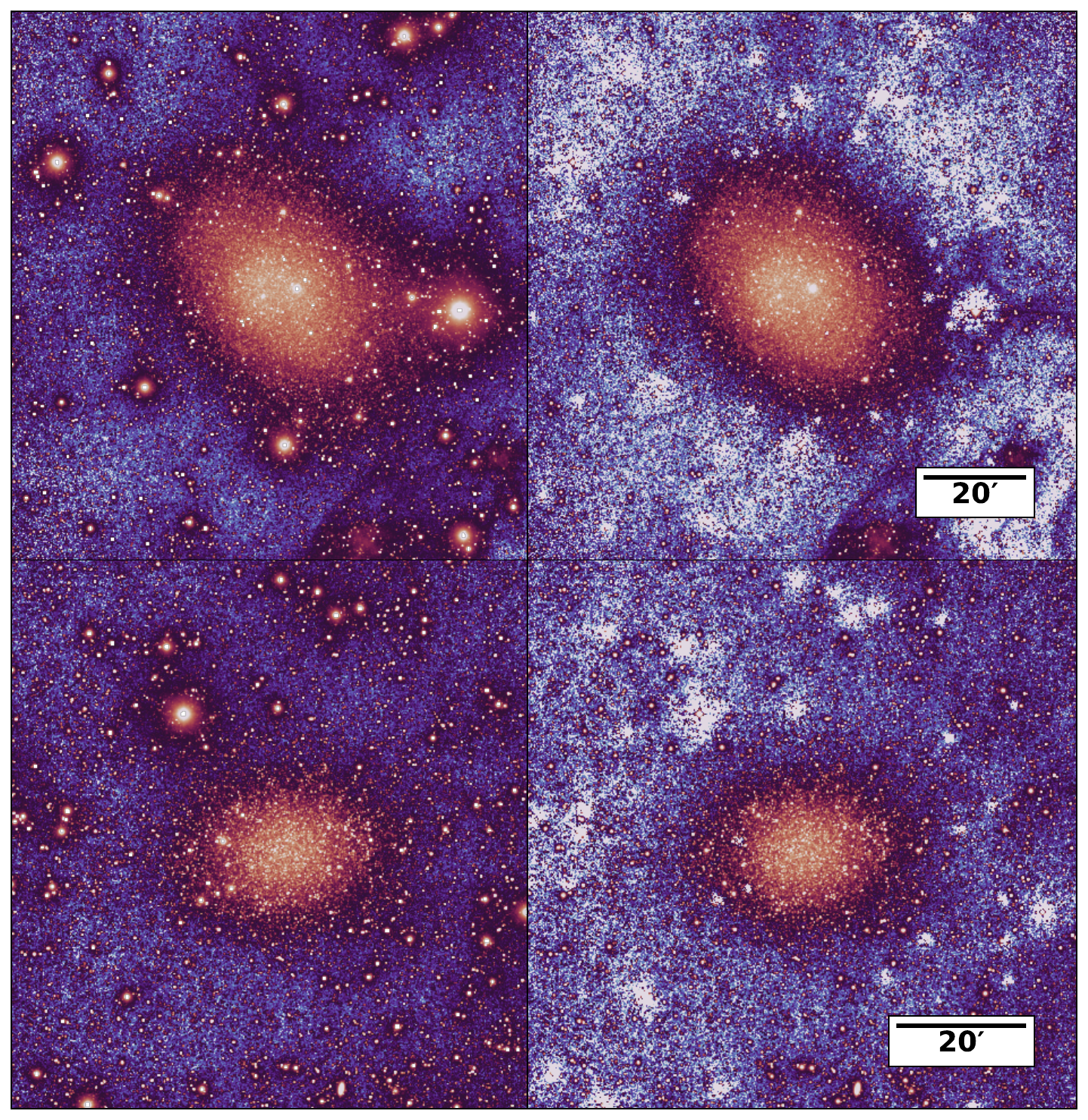}
    \caption{Mosaic showing the images of Fornax and Sculptor dwarf spheroidals before and after subtracting the scattered light of the brightest stars of the image (top and bottom, and left and right, respectively).} 
    \label{fig:starSubtractedComparison}
\end{figure}

Once the PSF was modelled, we subtracted the scattered light of all the stars of the images up to magnitude 12 in the G filter of Gaia (\citealt{gaia2023}; i.e. $m_{G} < 12\,\rm{mag}$) using the code MAHDI\footnote{https://github.com/CarlosMDLR/MAHDI} \citep[Mitigation Algorithm for Halo and Diffuse Illumination,][]{marreroDeLaRosa2026}. The images of Fornax and Sculptor before and after removing the stars are shown in Fig.~\ref{fig:starSubtractedComparison}.

\subsection{Masking}\label{sec:masks}

Masking the contaminants from the images is needed in order to confidently analyse and exploit the data. We have sequentially generated the masks, targeting one source of contamination at a time. The individual components that make up the final masks are the following:

\begin{itemize}
\item Galactic cirrus emission, which especially affects the field of Fornax. To identify them we took advantage of the 857 GHz maps from the Planck Legacy Archive \citep{planck2020}. A threshold of $1.5\ \rm{MJy/sr}$ was defined to reject regions where dust emission is strong enough to hamper the analysis.

\item Stars that do not belong to the target galaxy. We used Gaia DR3 \citep{gaia2023} to mask stars up to its limiting magnitude (i.e. $m_{G} \sim 21$). This effectively masks the central parts of the bright stars that have been subtracted---which tend to show residuals---and also masks faint non-subtracted stars.

\item M dwarfs from our Galaxy identified using Colour-Magnitude Diagrams (CMD) built from available catalogues. For the northern field (i.e. Leo~II) we used the SDSS DR16 catalogue \citep{sdssdr16}, and for the southern fields (i.e. Fornax and Sculptor) we used the Dark Energy Survey (DES) DR2 main catalogue \citep{Abbot2021}. The CMDs were built using the colour $(g-r)$, with the M dwarf population spanning roughly between $1.1 < (g-r) < 1.8$ (see e.g. \citealt{West2011}). The magnitude range in which we are able to identify these Galaxy stars is between 18 and 23 in the g-band.

\item Background galaxies were masked also using the DES DR2 main catalogue and the SDSS DR16 catalogue. We filtered the objects identified as galaxies, and masked them using the provided morphological parameters.
\end{itemize}

Once these four components were obtained, we combined them into a single mask. Manual masking of any remaining artefacts (e.g. residuals from the star subtraction) was performed in order to obtain the final masks, which are shown in App. \ref{App:Masks}.

\section{Analysis} \label{Sec:analysis}

The analysis performed on the data consists of extracting the radial distribution of light of the galaxies. These are also used to compute the structural parameters of the objects and their stellar masses.

\subsection{Surface brightness profiles extraction}
\label{sec:profileExtraction}

Extracting surface brightness profiles is a standard practice when analysing the integrated light of an object in the non-resolved regime, but it is uncommon for nearby systems, where the data are resolved or semi-resolved and stellar surface number density profiles can be obtained. In order to derive surface brightness profiles in semi-resolved data, we needed to change the methodology commonly used. The reason is that, in this regime, the number of stars in each resolution element is low enough to induce stochastic flux variations from pixel-to-pixel \citep{renzini1998, conroy2016, fastar2}, imposing constraints on how to extract the information.

First, we choose the statistical indicator to build the profiles. To find a representative value for a set of pixels (such as within an annulus), it is standard to use robust estimators: specifically, the median combined with sigma clipping (see e.g. \citealt{gnuastroRadialProf2024}). This approach is suitable for characterising smooth, diffuse systems, as it effectively filters out contaminants. Nevertheless, this methodology fails when applied in the semi-resolved regime. As a result of the stochastic fluctuations induced by the small number of stars in each resolution element, a sigma-clipping operation flags as outliers and rejects valid pixels from the target. For this reason, we build the profiles using a simple mean without applying any sigma clipping operation. Consequently, we ensure that we are not missing any signal from the galaxy, relying on the masks to filter out contamination (see Sec.~\ref{sec:masks}).

The adopted centre of an object is another critical parameter when extracting its radial light distribution, as any misplacement artificially flattens the innermost regions of the profile. Determining the centre is particularly challenging for dwarf spheroidal galaxies because their central regions are remarkably flat (see Fig.~\ref{fig:colourImages}), offering no clear brightness peak to anchor it. Furthermore, in the semi-resolved regime the importance of the centre is twofold due to pixel-to-pixel fluctuations. The amplitude of these fluctuations depends on the area covered by each annulus, meaning they progressively increase towards the inner regions where the sampled area decreases. In short, the innermost profile is not only sensitive to the accuracy of the adopted centre, but is also increasingly affected by stochasticity arising from the small number of stars sampled. Consequently, the measured profile inevitably depends on the  chosen centre.

Our approach to this issue is to work with a central region instead of with a specific centre, obtaining a family of surface brightness profiles. These profiles are later combined into a median profile and a set of envelopes characterising their distribution. This methodology removes any bias from any specific centre determination and effectively increases the number of stars used at each radial distance. The drawback of this approach is that the spatial resolution of the median profile is reduced (since we are combining measurements at slightly different radial distances) and the information of the explored region is washed out, as we are averaging the whole region.

For defining the region to explore, we compiled a set of centres published in the literature for each galaxy (details are given in App.~\ref{App:Centres}) and computed the mean and standard deviation of these coordinates. Then we draw random centres from this region, using a Monte Carlo approach and assuming a 2D Gaussian probability distribution. The central coordinates and the standard deviation of the exploration regions, together with the ellipticity and position angle (PA) used for the profiles (adopted from \citealt{munoz2018}) are listed in Table~\ref{tab:profileParams}.

\begin{table}[h]
\centering
\caption{Parameters used to derive the surface brightness profiles. The coordinates and the $\sigma$ (i.e. radius) define the exploration region from which the centres are drawn. The position angle is defined as the angle of the major axis measured from north to east.}
\begin{adjustbox}{width=\linewidth}
\begin{tabular}{c ccccc}
\toprule
Object & $\rm{RA}_{region}$ & $\rm{Dec}_{region}$ & $\rm{\sigma}_{region}$ & PA & Ellipticity \\
 & (deg) & (deg) & (arcsec) & (deg) &    \\
\midrule
Leo~II    &  $11^h13^m27.15^s$  & $+22^{\circ} 09^{\prime}09.0^{\prime\prime}$ & 10.0 & 38 & 0.07 \\
Sculptor &   $01^h00^m05.80^s$  & $-33^{\circ} 43^{\prime}08.2^{\prime\prime}$ & 16.3 & 92 & 0.33 \\ 
Fornax   &  $02^h39^m52.07^s$  & $-34^{\circ} 30^{\prime}40.2^{\prime\prime}$ & 12.5 & 45 & 0.29 \\
\bottomrule
\end{tabular}
\end{adjustbox}
\label{tab:profileParams}

\end{table}

\subsection{Surface brightness radial profiles}\label{Sec:sbprofiles}

\begin{figure*}[htb]
    \centering
    \includegraphics[width=\linewidth]{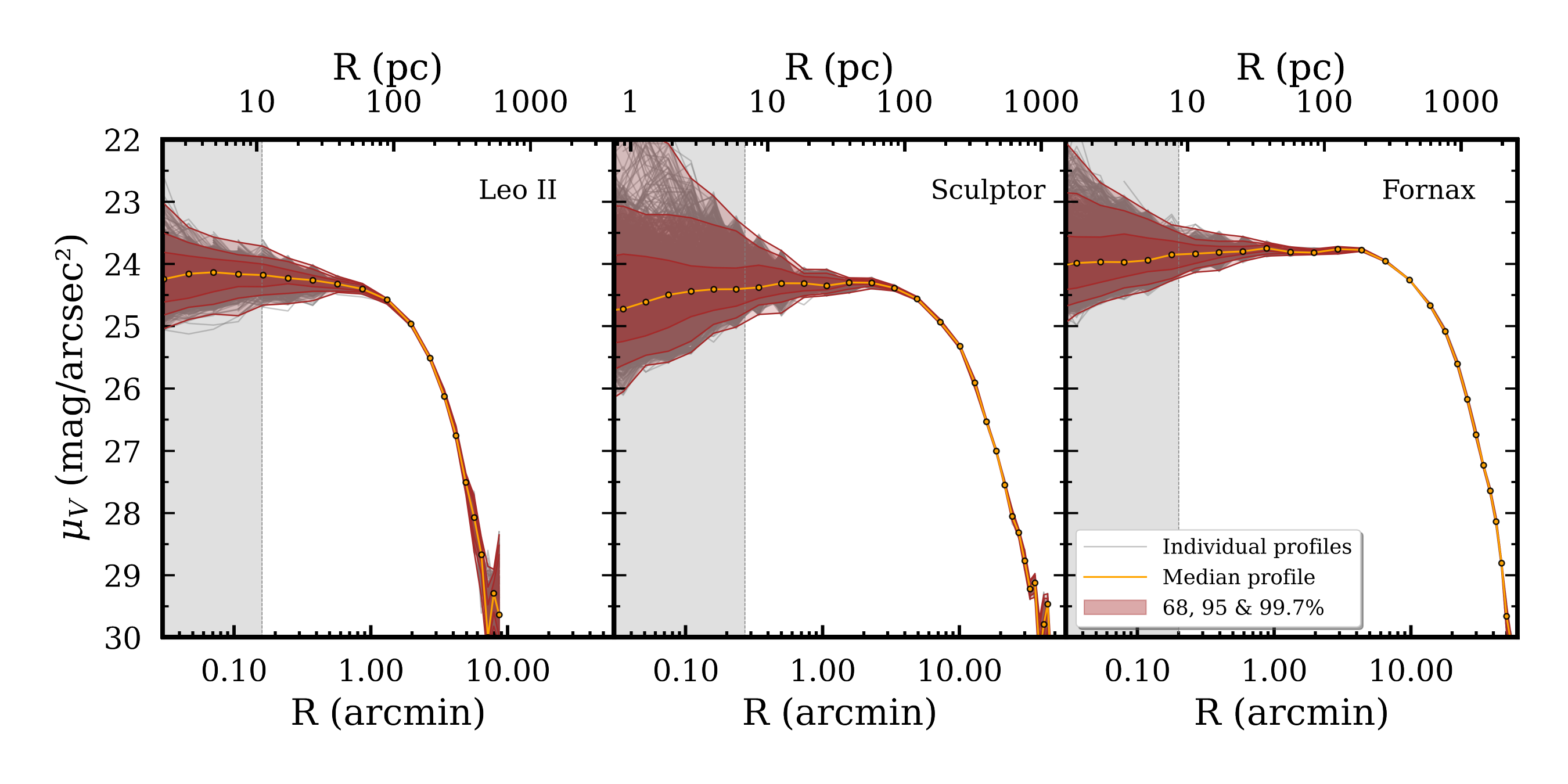}
    \caption{From left to right: radial surface brightness profiles of Leo~II, Sculptor, and Fornax in the V band. In grey lines the individual profiles obtained from the centre exploration, in yellow the median profile, and in red the contours enclosing 68, 95, and 99.7\% of the individual profiles. The shaded grey area corresponds to $1\sigma$ of the region in which the centre exploration is performed.} 
    \label{fig:lightProfiles}
\end{figure*}

The surface brightness profiles in the V-band for Leo~II, Sculptor, and Fornax are shown in Fig.~\ref{fig:lightProfiles}, with the corresponding tabulated values provided as supplementary material (Tables S1, S2, and S3). To convert from angular (arcmin) to physical (pc) units, we adopt the following distances: 233, 86, and 147 kpc for Leo~II, Sculptor, and Fornax respectively \citep{McConnachie2012}. A total of 2500 individual profiles are obtained with the Monte Carlo analysis, shown as the grey lines. The regions containing 68, 95, and 99.7\% of the profiles are shown in red, and the median profile is shown in yellow. The shaded grey area represents the radius $1\sigma$ of the region used to draw the centres of the profiles. The profiles are corrected from Galactic extinction, corresponding to the extinction of the Landolt V filter ($A_{\rm{Landolt\ V}}$): 0.046, 0.050, and 0.063 magnitudes for Leo~II, Sculptor, and Fornax respectively\footnote{\url{https://ned.ipac.caltech.edu/extinction_calculator}}. The individual profiles exhibit large variations in the innermost regions, driven by the stochasticity of the small number of stars in each annulus. Once combined, the fluctuations are reduced. 

\begin{figure}[h]
    \centering
    \includegraphics[width=0.9\columnwidth]{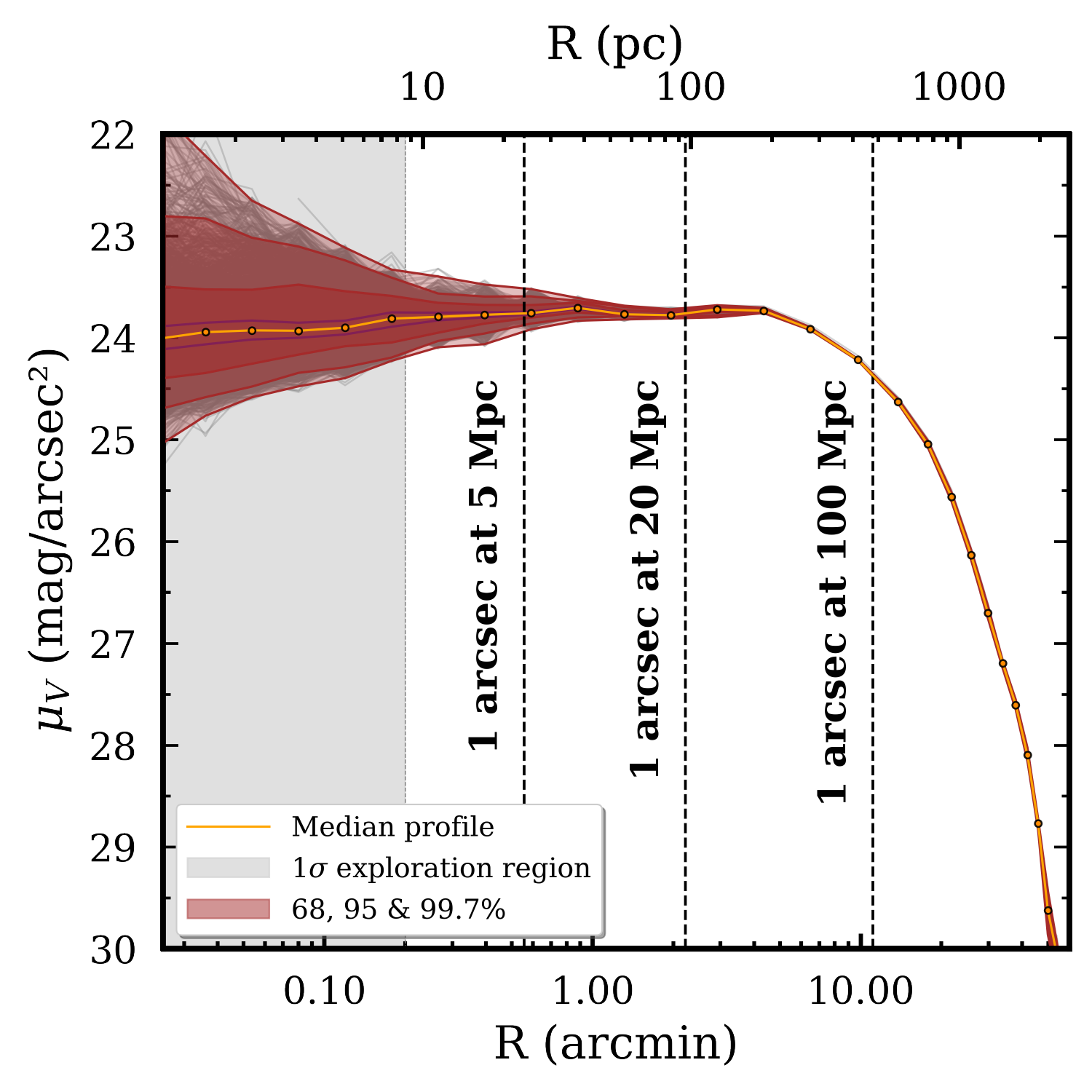}
    \caption{Surface brightness profile of Fornax similar to those shown in Fig.~\ref{fig:lightProfiles}. The regions that would be observable with a resolution of 1 arcsec if the galaxy were at different distances are those located towards the right of the dashed vertical lines.}
    \label{fig:fornaxResolutionExample}
\end{figure}

The next step is to model the profiles using a functional form. Historically, the \cite{sersic1968} function has been extensively and successfully used for fitting the profiles of galaxies, especially of early-type galaxies. This model reproduces nicely the outer parts of the profiles, but is incapable of capturing the inner flat distributions when these are present. This issue appears as we move to higher-resolution imaging and nearby objects. Fig.~\ref{fig:fornaxResolutionExample} shows the profile of Fornax, and the dashed vertical lines indicate the regions (to the right of each line) that would remain observable at 1\arcsec \ resolution if the galaxy were located at various distances. As can be seen, while at 100 Mpc the profile can be described by a S\'ersic, at closer distances a flat component is unveiled in the inner parts that cannot be described by a simple S\'ersic. To address this inability of the S\'ersic model to capture a power-law behaviour in the innermost region, we use the Core-S\'ersic function \citep{graham2003, trujillo2004}, which is designed to model these inner components. 

The Core-S\'ersic model combines an outer S\'ersic profile with an inner power-law with the following functional form:

\begin{equation} \label{eq:coreSersic}
    I(r) = I'[1+(r_{\rm{b}}/r)^{\alpha}]^{\gamma/\alpha}\ \exp \{-b[(r^{\alpha} + r_{\rm{b}}^{\alpha})/r_{\rm{e}}^{\alpha}]^{1/(n\alpha)}\},
\end{equation}
with

\begin{equation}
    I' = I_{\rm{b}}2^{-(\gamma/\alpha)}\ \exp\{b(2^{1/\alpha }r_{\rm{b}}/r_{\rm{e}})^{1/n}\},
\end{equation}

where $r_{\rm{b}}$ determines the change from the inner power-law to the outer S\'ersic, $\gamma$ is the internal slope, $\alpha$ controls the sharpness of the transition, $n$ is the S\'ersic index of the outer part, b is defined such that $r_{\rm{e}}$ characterises the effective radius of the galaxy, and $I_{\rm{b}}$ is the intensity at $r_{\rm{b}}$. We employed a Markov Chain Monte Carlo (MCMC) for the sampling, the details are given in App.~\ref{App:CoreSersicAppendix}. Table~\ref{tab:mcmcParams} compiles the resulting parameters for the median profiles assuming a Core-S\'ersic profile. Additionally, the inner slopes of the individual profiles are characterised and shown in App.~ \ref{App:InnerSlopes}.

\begin{table*}[h]
\centering
\caption{Median values obtained from modelling the observed median profiles with a Core-S\'ersic. The uncertainties of $r_{\rm{b}}$ and $r_{\rm{e}}$, when given in pc, incorporate the distance uncertainty.}
\begin{adjustbox}{width=0.8\linewidth}
\renewcommand{\arraystretch}{1.8}
\begin{tabular}{c cccccccc}
\toprule
Object & $\mu_{\rm{b,V}}$  & $\gamma$ & $\alpha$ & $r_{\rm{b}}$ & $r_{\rm{b}}$ & $r_{\rm{e}}$ & $r_{\rm{e}}$ & $n$\\
 & ($\rm{mag}/\rm{arcsec}^2$) &  &  & (arcmin) & (pc) & (arcmin) & (pc) & \\
\midrule
Leo~II & $25.08^{+0.40}_{-0.38}$ & $0.04^{+0.08}_{-0.09}$ & $3.2^{+1.2}_{-1.0}$ & $2.10^{+0.48}_{-0.49}$ & $143^{+33}_{-34}$ & $2.40^{+0.11}_{-0.10}$ & $162^{+12}_{-12}$ & $2.90^{+1.31}_{-1.26}$ \\
Sculptor & $24.76^{+0.33}_{-0.28}$ & $-0.09^{+0.05}_{-0.06}$ & $3.1^{+1.3}_{-1.1}$ & $6.58^{+1.48}_{-1.63}$ & $164^{+39}_{-42}$ & $10.97^{+0.54}_{-0.52}$ & $274^{+23}_{-23}$ & $3.27^{+1.12}_{-1.15}$ \\
Fornax & $23.84^{+0.11}_{-0.05}$ & $-0.06^{+0.02}_{-0.02}$ & $2.8^{+1.4}_{-1.2}$ & $3.89^{+1.67}_{-1.26}$ & $166^{+73}_{-56}$ & $15.83^{+0.18}_{-0.17}$ & $677^{+56}_{-56}$ & $0.78^{+0.03}_{-0.03}$ \\

\bottomrule
\end{tabular}
\end{adjustbox}
\label{tab:mcmcParams}
\end{table*}

\subsection{Stellar mass, effective radius, and core radius}\label{Sec:stellarMassDensity}

From the surface brightness profiles (Sec. \ref{Sec:sbprofiles}) we derive the main structural parameters, namely the effective radius, the core radius, and the total stellar mass of the galaxy.

The stellar mass density profiles are obtained by following the prescriptions given in \cite{Bakos2008} and these are integrated up to $\mu_{\rm{V}} = 29\ \rm{mag}/\rm{arcsec}^2$ to obtain the total stellar mass. The adopted value for the absolute magnitude of the Sun in the V-band ($m_{\rm{abs},\odot,V}$) is 4.78 mag, obtained from convolving the solar spectrum given by \cite{wilmer18} with the transmittance of the adopted V filter (see App. \ref{App:transmittances}). The conversion from surface brightness to stellar mass density additionally requires the mass-to-light (M/L) ratio. This M/L is frequently obtained by using colour information (e.g. \citealt{roediger2015}), but since the strength of the stochasticity depends on the filter (see App.~\ref{app:rgbProfiles}), the colour profile is not reliable and it is not feasible to follow this methodology. Instead, we make a rough estimation by assuming the age and metallicity of the dominant population of the galaxies and a Kroupa IMF \citep{kroupa2001} using the FASTAR package \citep{fastar1, fastar2}. Thus, we assume the following ages and metallicities: 9 Gyr and [Fe/H] =  -1.59 for Leo~II \citep{Mighell1996, kirby2011}, 10 Gyr and [Fe/H] =  -1.67 for Sculptor \citep{boer2011, kirby2011}, and 5.4 Gyr and [Fe/H] =  -1.01 for Fornax \citep{saviane2000, kirby2011}. The resulting stellar M/L are 1.29, 1.35, and 1.06 $\rm{M}_\odot/L_\odot$ respectively. The assumed distances are $233\pm14$ kpc for Leo~II, $86\pm6$ kpc for Sculptor, and $147\pm12$ kpc for Fornax \citep{McConnachie2012}. The total stellar masses obtained are $(0.85\pm0.10)\times10^6$ \Msun, $(1.51\pm0.21)\times10^6$ \Msun,\ and $(13.90\pm2.27)\times10^6$ \Msun. All three masses agree with those reported in \citet{hammer2018} within $1.3\sigma$, and with those in \citet{McConnachie2012} within $2.1\sigma$\footnote{An uncertainty of 14\%, corresponding to the average relative error of our own stellar mass estimates, has been assumed for the stellar mass values from \citet{McConnachie2012}.}.

We also compute the effective and the core radii directly from the profiles. The provided core radii correspond to the radii at which the flux is half the maximum flux of the profile. This definition for the core radius is adopted to deal with the inner decrease of light of the profiles (see Sec.~\ref{sec:discussionLimitations}). These parameters together with the aforementioned stellar masses are compiled in Table~\ref{tab:properties}. 

\begin{table*}[h]
\centering
\caption{Derived stellar masses, effective radii, and core radii. The provided uncertainties are based on the uncertainty in the median profile (App.~\ref{App:MedianUncertainty}) and, when involved, in the distance.}
\begin{adjustbox}{width=0.6\linewidth}
\begin{tabular}{c ccccc}
\toprule
Object & $M_*$ & $r_{\rm{e}}$ & $r_{\rm{e}}$ & $r_{\rm{c}}$ & $r_{\rm{c}}$ \\
  & ($10^6$ \Msun) & (arcmin) & (pc) & (arcmin) & (pc)   \\
\midrule

Leo~II&  $0.85\pm0.10$  & $2.39\pm0.01$ &     $162\pm10$ & $1.85\pm0.12$ & $126\pm11$ \\
Sculptor &  $1.51\pm0.21$  & $10.76\pm0.02$ &     $269\pm19$ & $8.20\pm0.20$ & $205\pm15$ \\
Fornax  &  $13.90\pm2.27$  & $15.60\pm0.02$ &     $667\pm54$ & $12.32\pm0.10$ & $527\pm43$ \\

\bottomrule
\end{tabular}
\end{adjustbox}
\label{tab:properties}
\end{table*}

\section{Discussion}\label{sec:Discussion}
\subsection{Comparison with stellar surface number density profiles}

In order to verify whether integrated light studies and star-counting studies agree on the inferred radial stellar distributions, we have compiled several stellar surface number density profiles from the literature and compared them with the surface brightness profiles obtained here. For this, we assumed a constant M/L (i.e. multiplying the whole stellar surface number density profile by a single value) for the star-counting profiles. We consider this to be the safest and least biased way of performing the comparison, since we do not need to make use of any extra information (such as number and distribution of different stellar populations). It is worth emphasising the physical meaning of the profiles: stellar surface number density profiles trace the spatial distribution of resolved stars and therefore sample only the brightest stellar populations. In contrast, surface brightness profiles trace the total flux emitted by the galaxy and thus reflect the contribution of its entire stellar content.

\begin{figure*}[h]
    \centering
    \includegraphics[width=\linewidth]{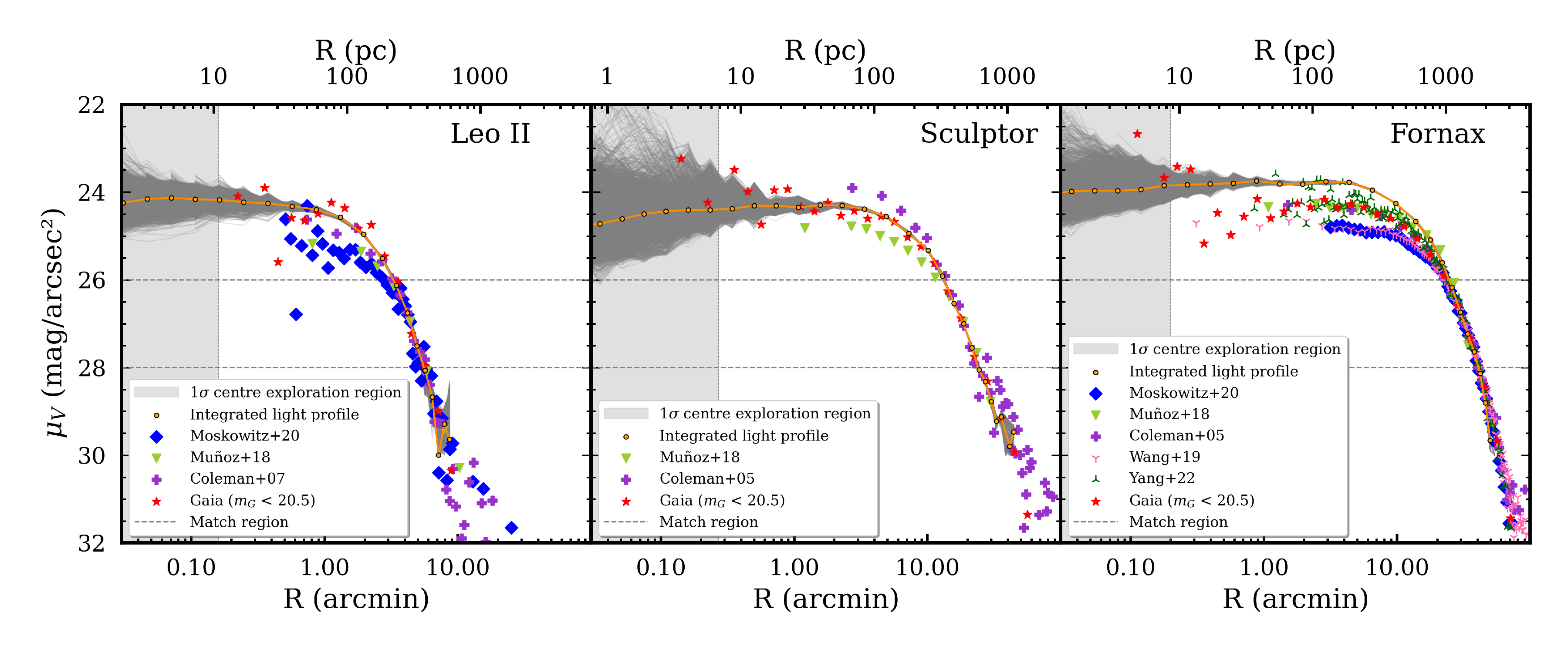}
    \caption{Comparison between the surface brightness profiles presented in this work and the stellar surface number density profiles from the literature. The former are shown in yellow lines, and the latter are shown as scattered data points. The horizontal dashed lines indicate the surface brightness range (from 26 to 28 $\rm{mag}/\rm{arcsec}^2$) at which the profiles have been matched assuming a constant M/L. The shaded grey region corresponds to $1\sigma$ of the region in which the centre exploration is performed. From left to right: Leo~II, Sculptor, and Fornax.} 
    \label{fig:starCountingComparison}
\end{figure*}

The comparison is shown in Fig~\ref{fig:starCountingComparison}. The surface brightness profiles obtained in Sec.~\ref{Sec:analysis} are shown in yellow, and the stellar surface number density profiles from the literature are shown in different colours and markers. The datasets have been matched over a radial range where both techniques are expected to be reliable. Star based methods suffer from incompleteness due to crowding when the stellar density is high enough (see e.g. \citealt{Dalcanton2012, wang2019, Ding2025}), causing the measured profile to be underestimated; on the other hand, integrated light methods are less reliable towards the outer regions, where they are sensitive to several issues (e.g. uncertainties in the background subtraction). Thus, we have selected an intermediate region in surface brightness (from 26 to 28 $\rm{mag}/\rm{arcsec}^2$) where we expect the impact of the aforementioned factors not to be severe. 

The number of compiled stellar surface number density profiles varies among the galaxies, with Fornax having the largest set of measurements. Most of the profiles were taken directly from published studies except the Gaia-based profiles \citep{gaia2023}, which were constructed from the probabilities of membership provided by \cite{battaglia2022}, selecting stars with $m_G < 20.5$ mag. The ellipticity and position angle used were the same as those given in Table~\ref{tab:profileParams}. The stellar surface number density profiles compiled come from: Leo~II \citep{coleman2007, munoz2018, moskowitz2020}, Sculptor \citep{coleman2005, munoz2018}, and Fornax \citep{coleman2005_fornax, munoz2018, moskowitz2020, wang2019, Yang2022}.

\subsubsection{Profile comparison}

The agreement between the profiles is good in the intermediate and outer regions, both across different studies and between the two methodologies. This is not trivial, since as previously stated the quantities measured by star-counting and integrated-light analyses are fundamentally different. The agreement, however, breaks down towards the innermost regions, where the profiles diverge and there is a mismatch between methods and even between studies. The most plausible explanation is stellar crowding (see e.g. \citealt{Dalcanton2012, wang2019, Ding2025}). Most profiles begin to diverge at approximately the same surface brightness, $\mu_{\rm{V}} \sim 26\ \rm{mag}/\rm{arcsec}^2$, supporting the interpretation that stellar crowding is the primary driver of the discrepancy since it depends on the projected stellar density. The comparison raises questions on the reliability of the inner regions of the stellar surface number density profiles, which systematically fall below the integrated-light profiles, suggesting that the stellar crowding might be more severe than expected (e.g. \citealt{wang2019} performed artificial star tests in Fornax and found severe crowding issues only at $r<10$ arcmin). As shown here, this crowding artificially distorts the profile, preventing an accurate characterisation of the central region.

\subsubsection{Radial range explored}

The radial ranges that can be reliably characterised are different and complementary between the two approaches. On the one hand, the star-counting method can trace the profile out to very large radii from the centre of the galaxy\footnote{For instance, \citealt{Yang2022} traced Fornax's radial stellar distribution out to $\sim 130\,\rm{arcmin}$.}, and it is ultimately limited by the ability to identify member stars at large distances from the centre of the galaxy. However, it suffers in the inner regions due to stellar crowding. On the other hand, integrated light profiles are not as reliable in the outer regions owing to several sources of uncertainty such as background-subtraction errors, imperfect masking of contaminants, and reduction artefacts. Nonetheless, it allows the profile to be traced much closer to the galaxy centre, since it is insensitive to crowding, and it is only limited by the stochasticity of the number of stars per resolution element.

Using integrated light techniques, we are able to trace the stellar profile to radii ten times smaller with respect to most previous works. 
Compared to number density profiles from ground-based imaging, we push the innermost data point---excluding the centre exploration region---inwards by up to an order of magnitude (as seen by comparing our Fornax profile with that of \citealt{moskowitz2020}).
In comparison with works using Gaia data (e.g. \citealt{Yang2022} or the profile obtained based on \citealt{battaglia2022}), the radial range covered beyond the uncertain exploration region is similar but with significant (i.e. 10 times) less scatter. 

\subsection{Limitations due to small number of stars} \label{sec:discussionLimitations}

The small number of stars within the innermost annuli produces stochastic fluctuations in the measurements of the surface brightness profiles. These regions have their stellar populations sampled by a low number of stars per resolution element, failing to fully sample the mass function. This regime is poorly understood, although some efforts have already been made to study it. For example, \cite{conroy2016} analysed HST observations of M~31 with low numbers of stars ($N_{\rm{stars}}$) per pixel, and \citealt{fastar2} provided stellar population models designed to address this semi-resolved regime.

In the inner regions, Leo~II profile is flat, but Sculptor and Fornax show a decrease of light in the inner parts (see Fig.~\ref{fig:lightProfiles} and Table~\ref{tab:mcmcParams}). In App.~\ref{App:Simulation} we show through simulations that this behaviour is consistent with the profiles having an underlying flat distribution but being affected by stochasticity. In addition, in App.~\ref{app:rgbProfiles} we explore how this stochasticity affects the surface brightness profile of Fornax in different filters, finding that the effect is stronger towards redder optical wavelengths. Therefore, the wavelength of the observations is important when analysing this semi-resolved regime. In these old systems ($\sim 10\ \rm{Gyr}$), bluer optical bands appear less susceptible to this stochastic effect and can provide more robust photometric measurements. This is because these bands are less affected by the scarcer bright Red Giant Branch (RGB) stars, which are the main contributors to the surface brightness fluctuations in the central areas of these galaxies.

\section{Conclusions}\label{sec:Conclusions}

In this work, we have obtained new data from three classical dwarf spheroidal galaxies (Leo~II, Sculptor, and Fornax) using small-aperture amateur telescopes. These data fall in the semi-resolved regime, where their stellar populations are partially resolved. As a result of their proximity, in the literature these systems are typically studied via star-counting. We instead adopted an integrated-light approach and compared the obtained surface brightness profiles with stellar surface number density profiles from previous works.

The surface brightness and stellar surface number density profiles are in good agreement within their respective limits. The number density profiles are limited by stellar crowding in the inner regions, and the surface brightness profiles are limited by background-subtraction uncertainties and contamination in the outer regions. The agreement in the intermediate regions found here is non-trivial, since the physical information that the profiles contain is not the same. The results of this comparison provide a bridge for integrating studies from dwarf galaxies across a large range of distance.

Integrated-light analyses of semi-resolved data, although an uncommon practice, are complementary to star-counting studies. This approach improves the characterisation of the inner regions of the stellar radial distributions in both radial extent---by a factor of ten---and shape. We identify the presence of stochastic effects in the innermost annuli of the profiles, where a low number of stars are found. In these regions, the stochastic fluctuations are dominated by the bright stars that contribute to the photometric measurements.

Analysing galaxies in the semi-resolved regime is expected to become increasingly common (based on facilities such as JWST, Euclid, or Roman) as resolution keeps improving and low-surface-brightness regions are probed, making it important to understand its intrinsic limitations. We have shown how the strength of the stochasticity on the surface brightness profiles depends on the wavelength. Thus, when performing photometry in this regime, bluer bands appear to provide more robust measurements and are therefore preferable for old systems when feasible.

Tables S1, S2, and S3 are only available in electronic form at the CDS via anonymous ftp to cdsarc.u-strasbg.fr (130.79.128.5) or via http://cdsweb.u-strasbg.fr/cgi-bin/qcat?J/A+A/.

\begin{acknowledgements}
We thank Giuseppina Battaglia and Jorge Sanchez Almeida for useful discussions. We thank the referee for the constructive comments that helped improve the quality of the manuscript. SGA and IRC acknowledge support from grant PID2022-140869NB-I00 from the Spanish Ministry of Science and Innovation. IT acknowledges support from the State Research Agency (AEI-MCINN) of the Spanish Ministry of Science and Innovation under the grant PID2022-140869NB-I00, financed by the Ministry of Science and Innovation, through the State Budget and by the Canary Islands Department of Economy, Knowledge, and Employment, through the Regional Budget of the Autonomous Community. MM acknowledges support from grant RYC2022-036949-I financed by the MICIU/AEI/10.13039/501100011033 and by ESF+, grant CNS2024-154592 financed by MICIU/AEI/10.13039/501100011033, and program Unidad de Excelencia Mar\'{i}a de Maeztu CEX2020-001058-M, financed by MCIN/AEI/10.13039/501100011033, and by the MaX-CSIC Excellence Award MaX4-SOMMA-ICE. CMR acknowledges financial support from the CoBEARD project (PID2021-128131NB-I00) and the Spanish Ministry of Science, Innovation, and Universities (MICIU) through PID2025-174546NB-I00 (TRAMWAYS project). This work is part of grant CEX2025-001609-S, awarded to the Instituto de Astrofísica de Canarias under the Severo Ochoa Centre of Excellence program and funded by MICIU/AEI/10.13039/501100011033. This research also acknowledges support from the European Union through the following grants: "UNDARK" and "Excellence in Galaxies - Twinning the IAC" of the EU Horizon Europe Widening Actions  programmes (project numbers 101159929 and 101158446). Funding for this work/research was provided by the European Union (MSCA EDUCADO, GA 101119830). Views and opinions expressed are however those of the author(s) only and do not necessarily reflect those of the European Union or European Research Executive Agency (REA). 

This work makes use of the following code:
\texttt{Gnuastro} \citep{gnuastro}, 
\texttt{Astrometry.net} \citep{Barron2008}, 
\texttt{photutils} \citep{larrybradley2023},
\texttt{SExtractor} \citep{BertinArnouts1996}, 
\texttt{numpy} \citep{harris2020array},
\texttt{scipy} \citep{2020SciPy-NMeth},
\texttt{SCAMP} \citep{Bertin2006}
\texttt{astropy} \citep{Astropy},

\end{acknowledgements}

\bibliography{bib}

\begin{appendix}

\section{Filter transmittances} \label{App:transmittances}

\begin{figure}[htbp]
    \centering
    \includegraphics[width=\linewidth]{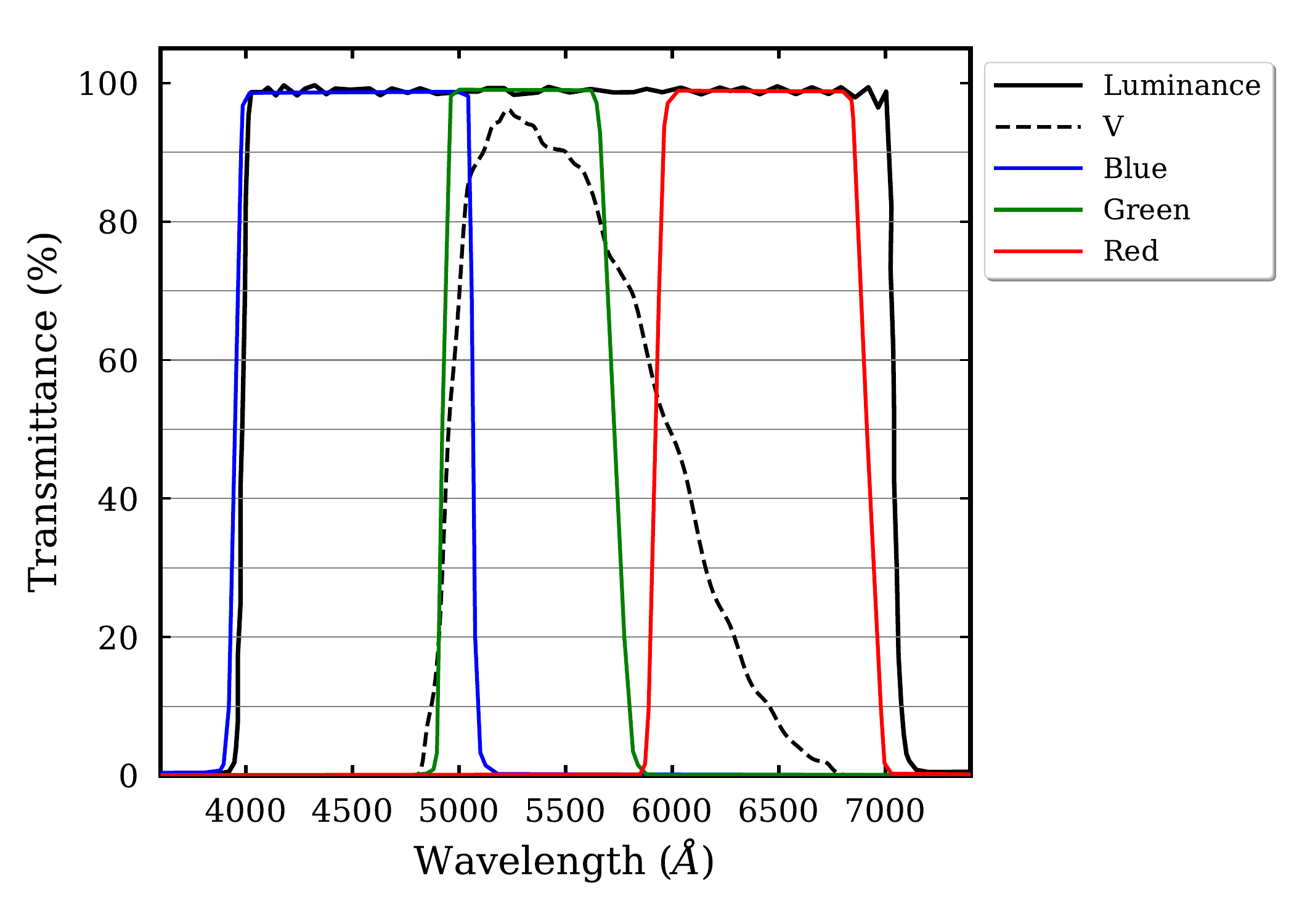}
    \caption{Transmittances of the five commercial filters used in this study. The four filters used for the observations are the \textit{Astrodon Generation 2 I-Series Filters} and the V filter used to calibrate the luminance measurements is the \textit{UBVRI Bessell V-Filter}.} 
    \label{fig:transmittances}
\end{figure}

The filters used for the observations are the \textit{Astrodon Generation 2 I-Series Filters}, and the V filter onto which the luminance measurements are transformed is the commercial \textit{UBVRI Bessell V-Filter}. The transmittances are shown in Fig.~\ref{fig:transmittances}. The factor adopted for transforming the Luminance filter to the standard V is an average value obtained from convolving spectra from Gaia stars with both filters and comparing the resulting magnitudes. The resulting factor is $m_{\rm{V}} = m_{\rm{lum}}-0.04$.

\section{Masks used in the analysis} \label{App:Masks}
\begin{figure*}[h]
    \centering
    \includegraphics[width=0.87\linewidth]{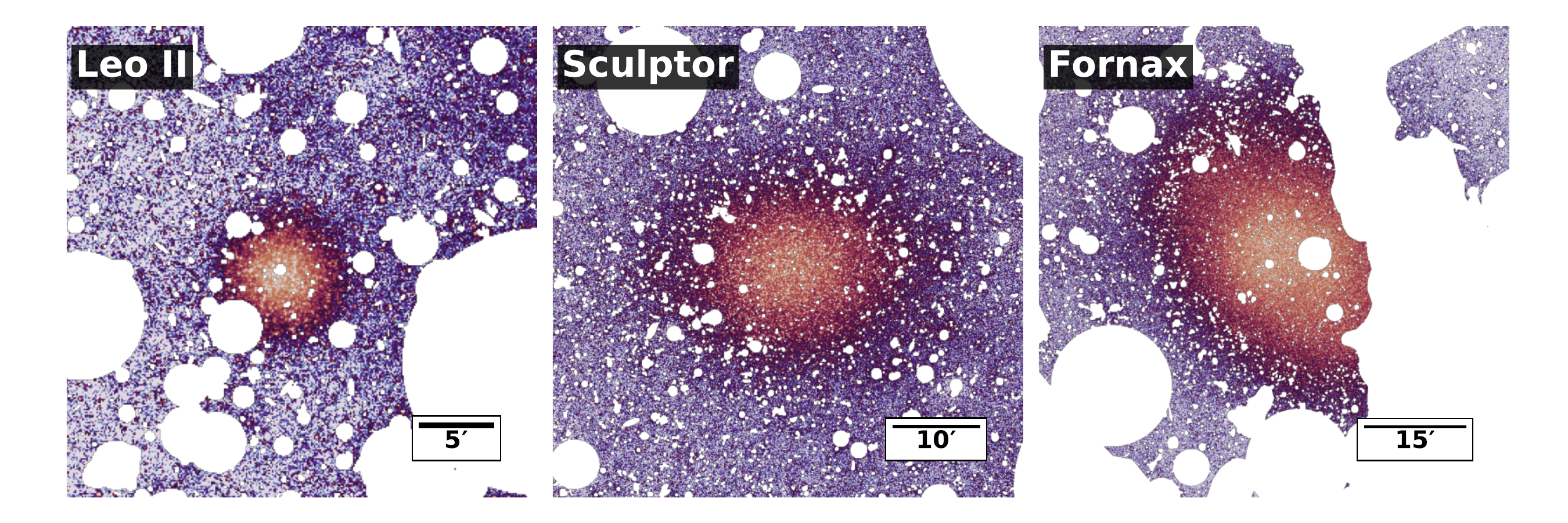}
    \caption{From left to right: images of Leo~II, Sculptor, and Fornax after the masking process described in Sec.~\ref{sec:masks}.} 
    \label{fig:masks}
\end{figure*}

Fig. \ref{fig:masks} shows the images of Leo II, Sculptor, and Fornax after applying the masks adopted for the analysis. The masks were obtained following the procedure described in Sec. \ref{sec:masks}. We note that the right-hand side of the Fornax dwarf spheroidal is affected by a column of Galactic cirrus emission, which is masked accordingly.

\section{Definition of the centre exploration region} \label{App:Centres}

To define a region over which the centre is explored, we compiled a set of the most recent centre estimates reported in the literature. Based on them we have defined a circular exploration region for building the surface brightness profiles. The centre of our exploration region is defined as the mean of the compiled centres, then the standard deviation in RA and Dec is obtained, and the largest is used as the radius of the circular region to explore. The exception has been Leo~II, where the majority of the reported centres are clearly off from our imaging; for this galaxy we have also taken the standard deviation of the centres for the size of the region, but assumed the reported centre with better visual agreement with our data (i.e. the one by \citealt{munoz2018}). The compiled centres are given in Table~\ref{tab:FornaxCentres}.

\begin{table}[h]
\centering
\caption{Summary of the compiled centres from the literature. These have been used to define the exploration region used in the characterisation of the surface brightness profiles. }
\label{tab:FornaxCentres}
\small
\resizebox{\columnwidth}{!}{
\begin{tabular}{cccc}
\toprule
Galaxy & RA (J2000) & Dec (J2000) & Source \\

\midrule

Leo~II & $11^h13^m27.05^s$  & $+22^{\circ} 09^{\prime}10.4^{\prime\prime}$ & \cite{munoz2018} \\
& $11^h13^m25.51^s$  & $+22^{\circ} 09^{\prime}02.5^{\prime\prime}$ & \cite{Moneli2025} \\
& $11^h13^m28.80^s$  & $+22^{\circ} 09^{\prime}06.0^{\prime\prime}$ & \cite{coleman2007} \\
& $11^h13^m29.00^s$  & $+22^{\circ} 09^{\prime}12.0^{\prime\prime}$ & \cite{Mateo1998} \\
& & & \\

Sculptor & $1^h00^m04.40^s$  & $-33^{\circ} 43^{\prime}07.0^{\prime\prime}$ & \cite{munoz2018} \\
& $1^h00^m00.70^s$  & $-33^{\circ} 43^{\prime}07.0^{\prime\prime}$ & \cite{ArroyoPolonio2024}\\

& & & \\

Fornax & $2^h39^m50.90^s$  & $-34^{\circ} 30^{\prime}53.0^{\prime\prime}$ & \cite{Yang2022} \\
& $2^h39^m50.90^s$  & $-34^{\circ} 30^{\prime}54.0^{\prime\prime}$ & \\
& $2^h39^m51.00^s$  & $-34^{\circ} 30^{\prime}49.0^{\prime\prime}$ & \\
& $2^h39^m53.00^s$  & $-34^{\circ} 30^{\prime}32.0^{\prime\prime}$ & \cite{wang2019} \\
& $2^h39^m53.00^s$  & $-34^{\circ} 30^{\prime}21.0^{\prime\prime}$ & \\
& $2^h39^m53.00^s$  & $-34^{\circ} 30^{\prime}25.0^{\prime\prime}$ & \\
& $2^h39^m53.00^s$  & $-34^{\circ} 30^{\prime}32.0^{\prime\prime}$ & \\
\bottomrule
\end{tabular}
}
\end{table}

\section{Characterising surface brightness radial profiles} \label{App:CharacteriseProfs}

\subsection{Median profile uncertainty}\label{App:MedianUncertainty}

\begin{figure*}[h]
    \centering
    \includegraphics[width=0.88\linewidth]{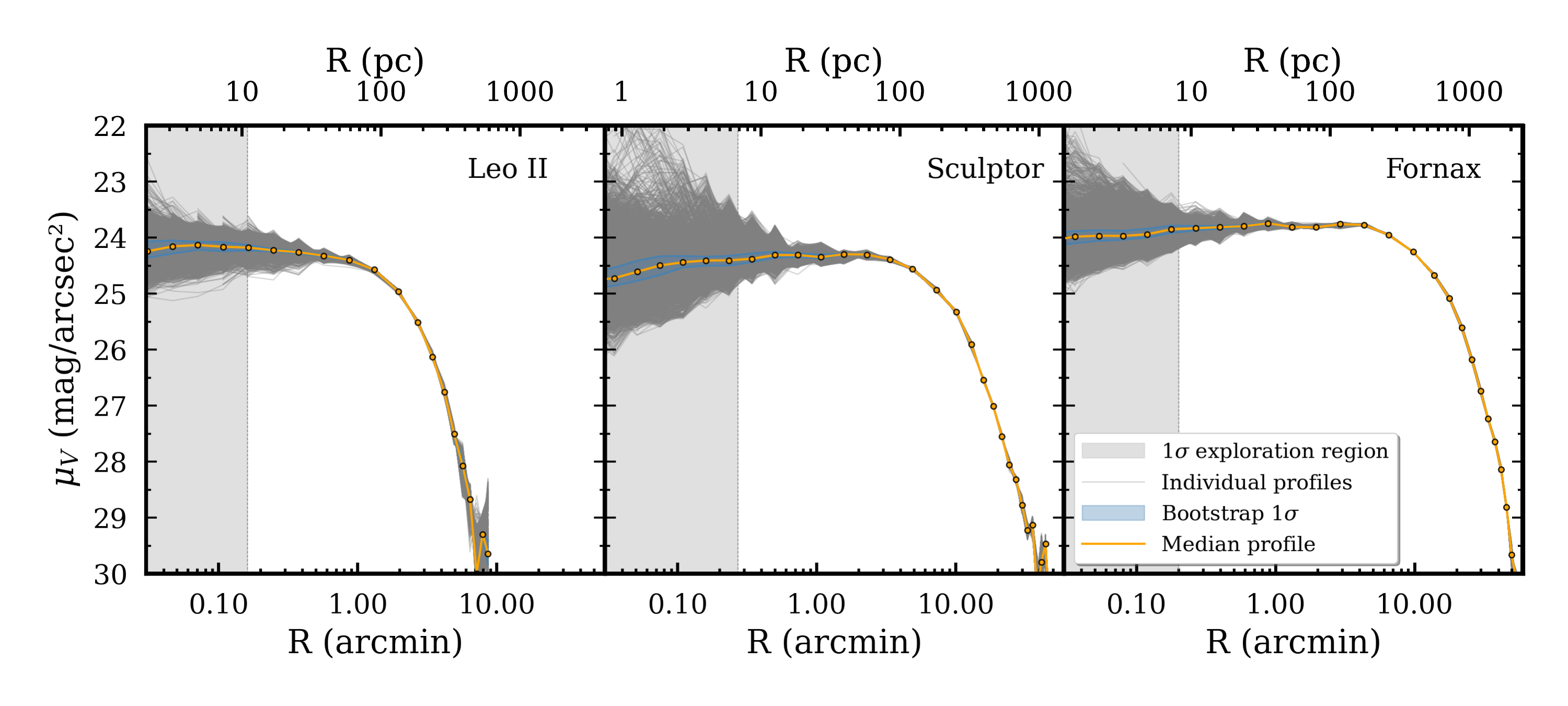}
    \caption{From left to right: radial surface brightness profiles of Leo~II, Sculptor, and Fornax. The individual profiles obtained from the centre exploration are represented in grey and the median profiles in yellow. The shaded grey region corresponds to $1\sigma$ of the region in which the centre exploration is performed. The blue region shows the $1\sigma$ uncertainty resulting from bootstrapping the individual profiles.} 
    \label{fig:lightProfilesBootstrap}
\end{figure*}

The diversity of profiles shown in Fig.~\ref{fig:lightProfiles} has its origin in the stochasticity of the innermost annuli, but it does not represent the uncertainty in measuring the median profile. To estimate this uncertainty, we use a bootstrap approach: we iteratively combine random subsets of the individual profiles and take the variation in the resulting distribution as the uncertainty in the median profile. The median profiles with their $1\sigma$ uncertainty computed after generating 500 median profiles using subsets of 25 profiles are shown in Fig.~\ref{fig:lightProfilesBootstrap}. 

\subsection{Core-S\'ersic modelling} \label{App:CoreSersicAppendix}

Modelling the median surface brightness profiles by assuming a Core-S\'ersic has been performed using a MCMC algorithm, implemented through the python package \texttt{emcee} \citep{Foreman2013}. The exploration is performed in $\rm{mag}/\rm{arcsec}^2$, the log-likelihood being obtained from the analytical expression of the core-S\'ersic (Eq. \ref{eq:coreSersic}) after it being converted into magnitudes. Since the characterisation performed in App.~\ref{App:MedianUncertainty} accounts for the centre exploration, the uncertainties in the outer regions are not captured. To deal with this, we allow the MCMC to explore a minimum common error for the data points ($\sigma_{\rm{floor}}$), thus avoiding the MCMC to over fit the outer region. For the exploration, we used 16 walkers, 5000 steps, a burn-in of 1000 steps, and flat priors (summary in Table~\ref{tab:Priors}).  The log-likelihood is:

\begin{equation}
    \ln \mathcal{L}(\theta) = -\frac{1}{2}\sum_{i=1}^{N}\left[{ \frac{(y_{i,\rm{obs}} - f(R_i|\theta))^2}{\sigma_{\rm{eff,i}(\theta)}^2} + \rm{ln}(2\pi \sigma_{eff,i}^2(\theta)})\right],
\end{equation}

where

\begin{equation}
    \sigma_{\rm{eff,i}} = \sqrt{\sigma_i^2 + \sigma^2_{\rm{floor}}},
\end{equation}

and $f(R|\theta)$ being the Core-S\'ersic intensity transformed into magnitudes. \\

\begin{table}[h]
\centering
\caption{Summary of the flat priors used for the MCMC.}
\label{tab:Priors}
\small
\setlength{\tabcolsep}{20pt} 
\begin{tabular}{cc}
\toprule
Parameter & Prior \\
\midrule
$\mu_b$ ($\rm{mag}/\rm{arcsec}^2$)  & Free  \\
$\gamma$  & -1.0 < $\gamma$ < 1.0 \\
$\alpha$  & 0.0 < $\alpha$ < 5.0 \\
$r_{\rm{b}}$ (arcmin)  & 0.1 < $r_{\rm{b}}$ < 10.0\\
$r_{\rm{e}}$ (arcmin) & 0.0 < $r_{\rm{e}}$ < 50.0 \\
$n$  & 0.01 < $n$ < 5 \\
$\sigma_{\rm{floor}}$ ($\rm{mag}/\rm{arcsec}^2$) & 0.01 < $\sigma_{\rm{floor}}$ < 0.5 \\
\bottomrule
\end{tabular}
\end{table}

The corner plots are shown in the top panels of Figs.~\ref{fig:LeoII_full}, \ref{fig:Sculptor_full}, and ~\ref{fig:Fornax_full}, and the comparisons between the observed profiles and the best Core-S\'ersic models found are shown in the bottom panels of the same figures, for Leo~II, Sculptor, and Fornax, respectively. A compilation of the median parameter values is given in Table~\ref{tab:mcmcParams}.

The corner plots show a fair amount of degeneracy between  different parameters (e.g. $r_{\rm{e}}$ with $r_{\rm{b}}$ or $r_{\rm{e}}$ with $n$). Nonetheless, the parameter $\gamma$, which is the most relevant parameter concerning the inner region of the profiles, is well constrained and non-degenerate. It is worth noting that $\sigma_{\rm{floor}}$ is, as expected, independent.

\subsection{Distribution of inner slopes}\label{App:InnerSlopes}

\setcounter{figure}{4}

\begin{figure*}[htbp]
    \centering
    \includegraphics[width=0.9\linewidth]{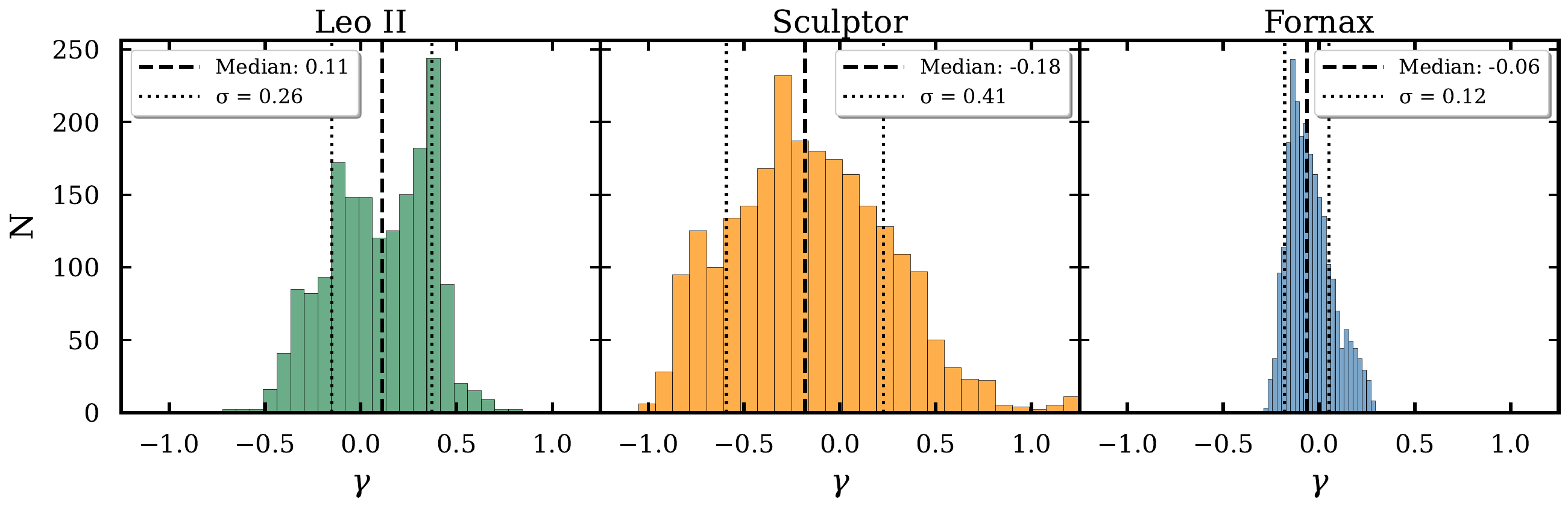}
    \caption{From left to right: inner slope distributions of Leo~II, Sculptor, and Fornax. The mean value and the standard deviation of the distributions are shown in dashed and dotted lines respectively. The widths of the bins have been calculated using the Freedman–Diaconis rule \citep{Freedman1981}. $\gamma$ values have been determined using the regions of the profiles inside 0.4, 1, and 1.5 arcmin for Leo~II, Sculptor, and Fornax respectively.} 
    \label{App:innerSlopesDistributions}
\end{figure*}

Apart from characterising the median profile, we study the distribution of inner slopes that the individual profiles exhibit. For each of the profiles, we fit a power law to the inner region to the radius where the transition to the outer part begins (i.e. the cored component), which corresponds approximately to r < 0.4, 1.0, 1.5 arcmin for Leo~II, Sculptor, and Fornax, respectively. Note that the profiles with bins containing large fluctuations due to the small number of stars show irregular inner profiles, which are not well captured by a power law. Nonetheless, the whole distribution of slopes is useful in order to characterise the diversity of profiles. The resulting distributions are shown in Fig.~\ref{App:innerSlopesDistributions}. The narrowest distribution is found for Fornax, having every gamma value well within -0.25 < $\gamma$ < 0.25, and the widest that of Sculptor, showing a range of -1.0 < $\gamma$ < 1.0. The median inner slopes (and $1\sigma$ dispersions) are 0.11 (0.26), -0.18 (0.41), and -0.06 (0.12) for Leo~II, Sculptor, and Fornax, respectively.

\section{Tests of stochastic effects due to small numbers of stars}\label{App:StatisticalInnerBehaviour}

\subsection{Surface brightness profiles of Fornax at different wavelengths} \label{app:rgbProfiles}

Analysing surface brightness profiles in different bands can provide insight into how stochasticity in photometric measurements varies with wavelength when only a few stars per resolution element are available. As discussed in Sec. \ref{sec:discussionLimitations}, not having enough stars results in an incomplete sampling of the mass function, and the strength of this effect will potentially vary depending on the wavelength. This occurs because sampling incompletely the mass function leads to an uneven representation of different stellar populations, each contributing differently to the total light budget (e.g. RGB stars and main sequence stars). Since these populations differ in brightness and colour, the impact of this stochasticity is inherently wavelength-dependent: bands where intrinsically luminous, but rare, populations dominate will be more strongly affected. This wavelength dependence is consistent with results from \cite{fastar2}, who investigated the effect of $N_{\rm{stars}}$ on the measured $(g-r)$ colour of stellar population models, reporting that the measured colour tends to become bluer as the number of stars decreases.

\begin{figure}[htbp]
    \centering
\includegraphics[width=\columnwidth]{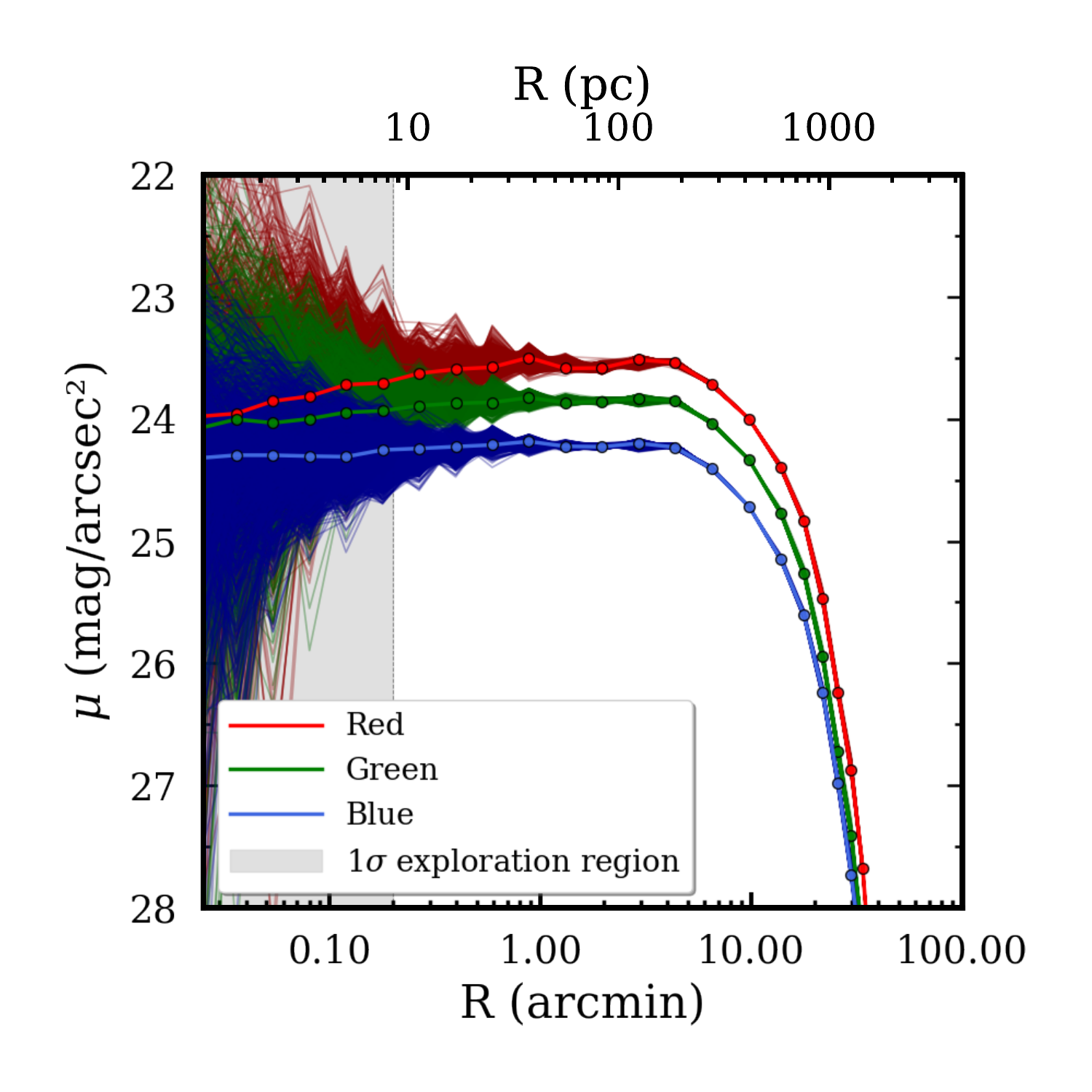}
    \caption{Observed surface brightness profiles of Fornax in the Blue, Green, and Red bands. The individual profiles obtained from the exploration of centres are shown in darker thinner lines, and the median profiles are shown in brighter, thicker lines. The $1\sigma$ of the region in which the centre exploration is performed is shown in grey. The stochasticity accentuates as moving towards redder bands, this is a direct result of the very few bright Red Giant Stars in the central regions.} 
    \label{fig:rgb}
\end{figure}

We have additional observations in three filters (Blue, Green, and Red, see App.~\ref{App:transmittances}) that, albeit shallower than the luminance data, can be used to test this expectation. Fig.~\ref{fig:rgb} shows the Blue, Green, and Red profiles of Fornax after performing 2500 iterations exploring different centres. The outer regions are not reliable due to the shallowness of these data compared to the luminance imaging, but the inner regions ($R \le 10$ arcmin) show a clear trend. For an old system like Fornax, the low number of bright red stars in the inner regions produces a deficiency of light which is accentuated towards the redder bands. As a result, the measured colour becomes bluer towards the centre. This appears to be consistent with the physical interpretation previously suggested, since redder bands are increasingly sensitive to RGB stars. 

\subsection{Simulating Fornax's inner deficiency of light}
\label{App:Simulation}

\begin{figure*}[htbp]
    \centering
    \includegraphics[width=0.8\linewidth]{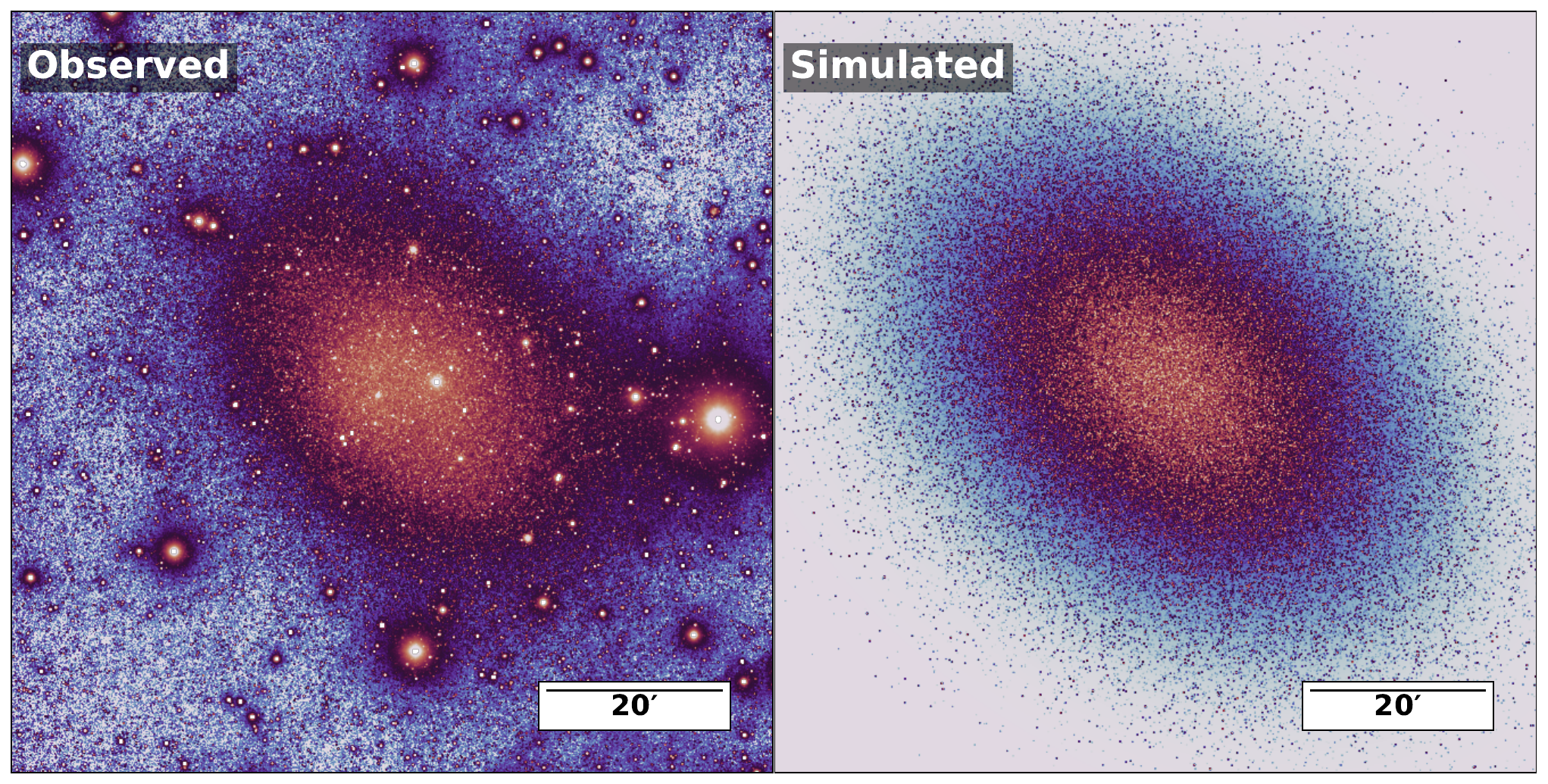}
    \caption{Left: Observed Fornax dwarf spheroidal galaxy. Right: Simulated galaxy assuming the best Core-S\'ersic fit but with a flat inner region (i.e. $\gamma = 0$). Both panels share the same logarithmic stretch and colour scale.} 
    \label{app:observedVsSimulated}
\end{figure*}

To further test whether the inner regions of the surface brightness profiles are affected by stochastic effects of the brightest stars, we perform a simulation of Fornax in which we assume an intrinsic profile for the galaxy, allowing us to assess how well the input profile is recovered. A genuine central light deficiency is difficult to interpret from a physical standpoint (see e.g. \citealt{trujillo2001}); we therefore assume an intrinsic flat core and test whether the deficiency observed in Fornax can be reproduced as a consequence of stochastic effects due to a small number of bright stars in the inner regions.

We used a custom python code to build a realistic mock image of the Fornax dwarf spheroidal galaxy, mimicking the properties of our data (namely the FOV and pixel scale). Since the data are partially resolved, we define two distinct components: a diffuse component, accounting for the unresolved populations of the galaxy, and a resolved component representing the brighter population that appears resolved in our data. The unresolved component is modelled with a Core-S\'ersic with the parameters listed in Table~\ref{tab:mcmcParams}, but with a flat inner region (i.e. $\gamma = 0$). This same Core-S\'ersic model is also used as a probability density function (PDF) for generating the resolved component.
The number and magnitudes of the stars of the resolved component come from the Fornax CMD obtained from the DES DR2 main catalogue (see Sec. \ref{sec:masks}), preserving only the bona fide stars that belong to Fornax. After this, these stars were placed in the simulation by randomly sampling the generative PDF. The flux of both components (diffuse and resolved) of the simulated galaxy is matched to the observed profile at a radius where the individual profiles exhibit little variation. In Fig.~\ref{app:observedVsSimulated} a visual comparison between the observed Fornax dwarf spheroidal galaxy and our simulation is shown.

\begin{figure}[htbp]
    \centering
\includegraphics[width=\columnwidth]{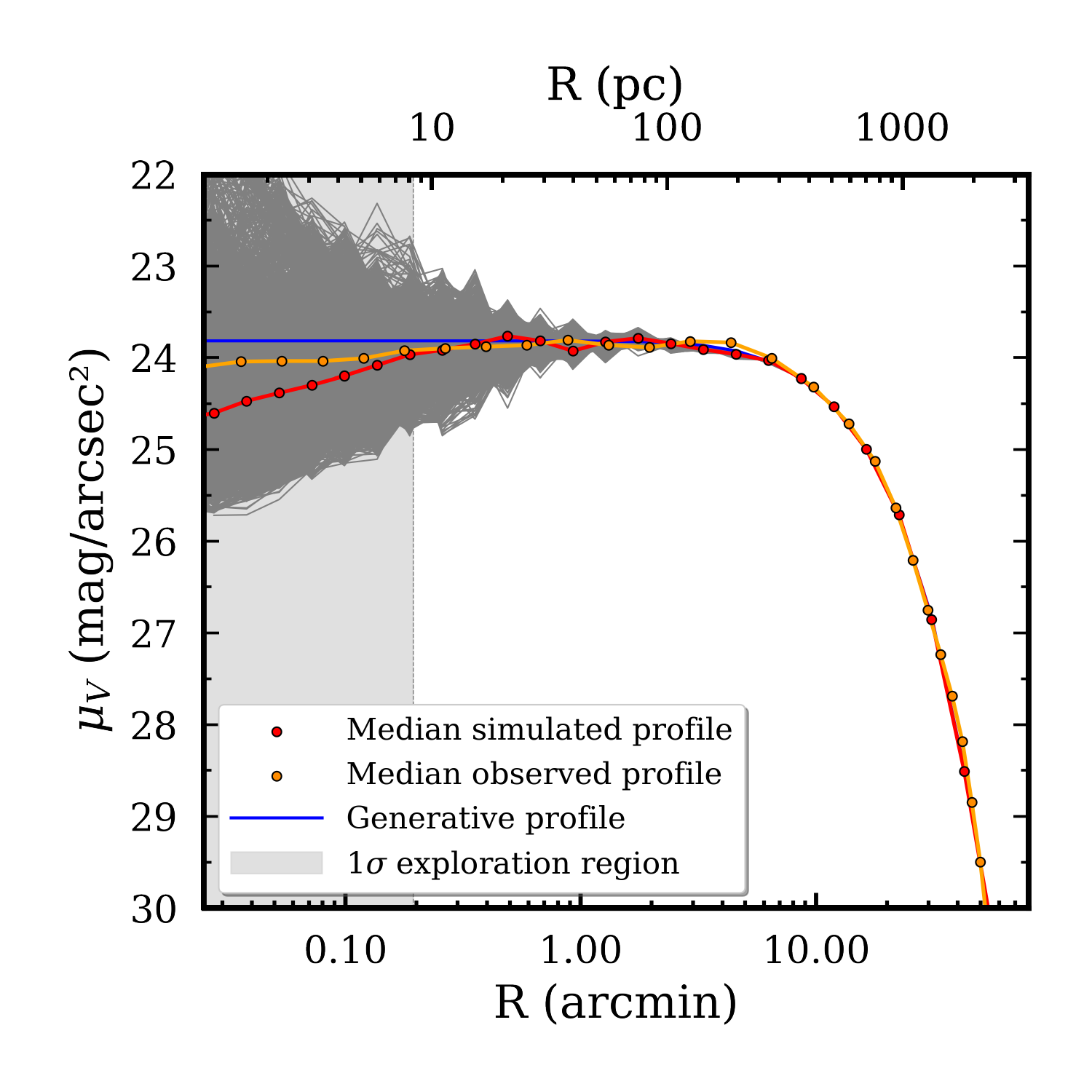}
    \caption{Surface brightness profiles obtained from simulating the Fornax dwarf spheroidal galaxy. The blue profile shows the simulated profile, the yellow profile shows the observed one, and the grey and red lines show the individual profiles from the centre exploration and their median, respectively. The median surface brightness profile of the simulated galaxy, despite being generated from an intrinsically flat profile, exhibits an inner deficiency (red profile).} 
    \label{fig:centralDeficiencySim}
\end{figure}

We, then, derive the surface brightness profile of this mock image as described in Sec.~\ref{sec:profileExtraction}. Fig.~\ref{fig:centralDeficiencySim} shows the profiles obtained, where the assumed intrinsic profile (blue) alongside the observed (yellow) and simulated profiles (grey and red for the median) can be compared. The simulation, despite being generated from a flat profile, exhibits an inner deficiency (i.e. negative $\gamma$ slopes). This shows how the shapes of the inner profiles are modified, consistent with stochastic effects arising from a few bright stars within each resolution element.

\begingroup
\setcounter{section}{4}
\setcounter{figure}{1}
\renewcommand{\thefigure}{\Alph{section}.\arabic{figure}}
\endgroup

\begin{figure*}[h]
    \centering
    \begin{subfigure}{\linewidth}
        \centering
        \includegraphics[width=0.8\linewidth]{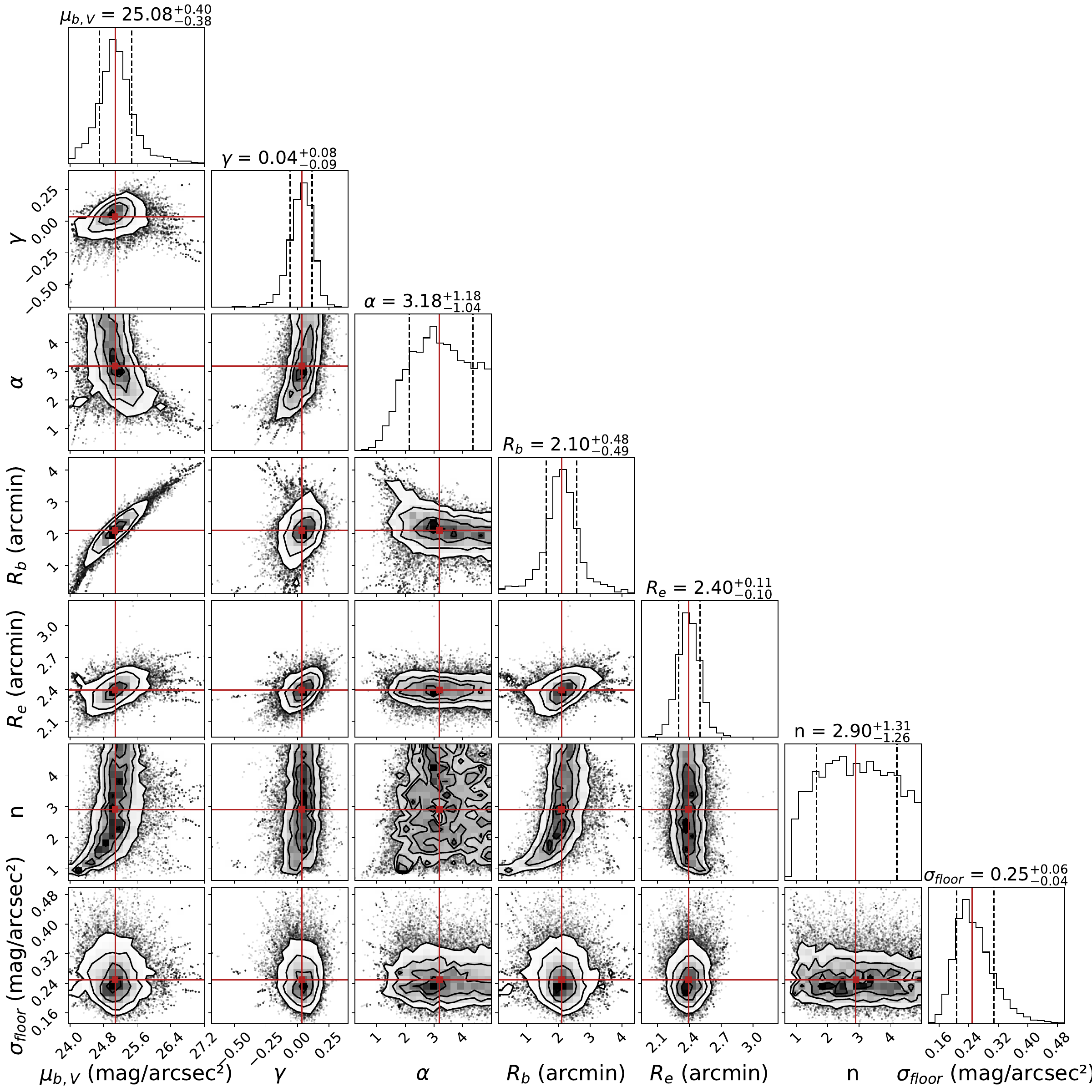}
    \end{subfigure}

    \vspace{0.3cm}
    
    \begin{subfigure}{\linewidth}
        \centering
        \includegraphics[width=0.5\linewidth]{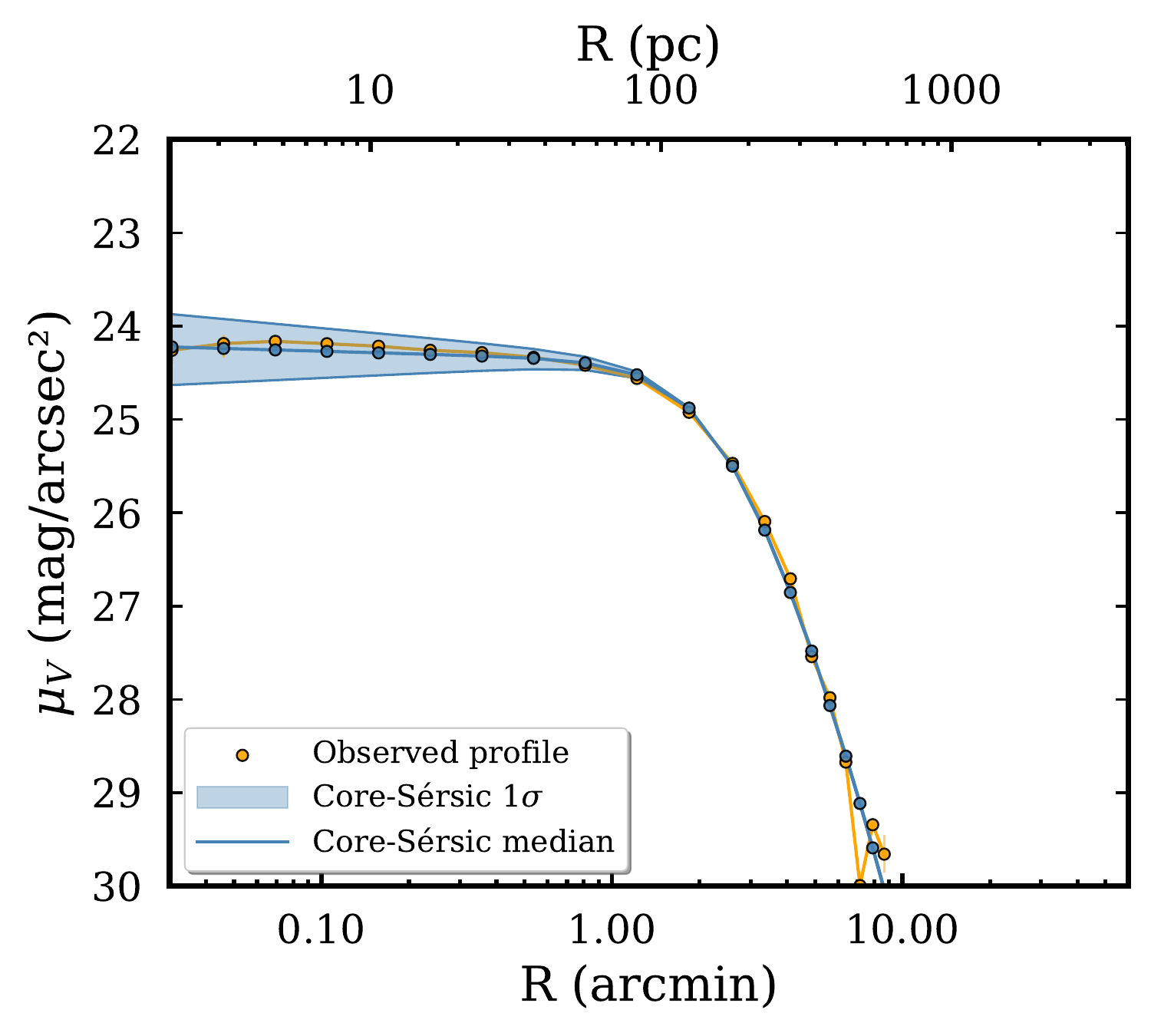}
    \end{subfigure}
    \caption{Core-S\'ersic modelling of the median surface brightness profile of Leo~II. Top: posterior distributions of the Core-S\'ersic parameters. Bottom: comparison between the observed profile and the best-fit model obtained. The observed profile is shown in yellow, and the best-fit model inferred from the MCMC is shown in blue.}
    \label{fig:LeoII_full}
\end{figure*}

\begin{figure*}[h]
    \centering
    \begin{subfigure}{\linewidth}
        \centering
        \includegraphics[width=0.8\linewidth]{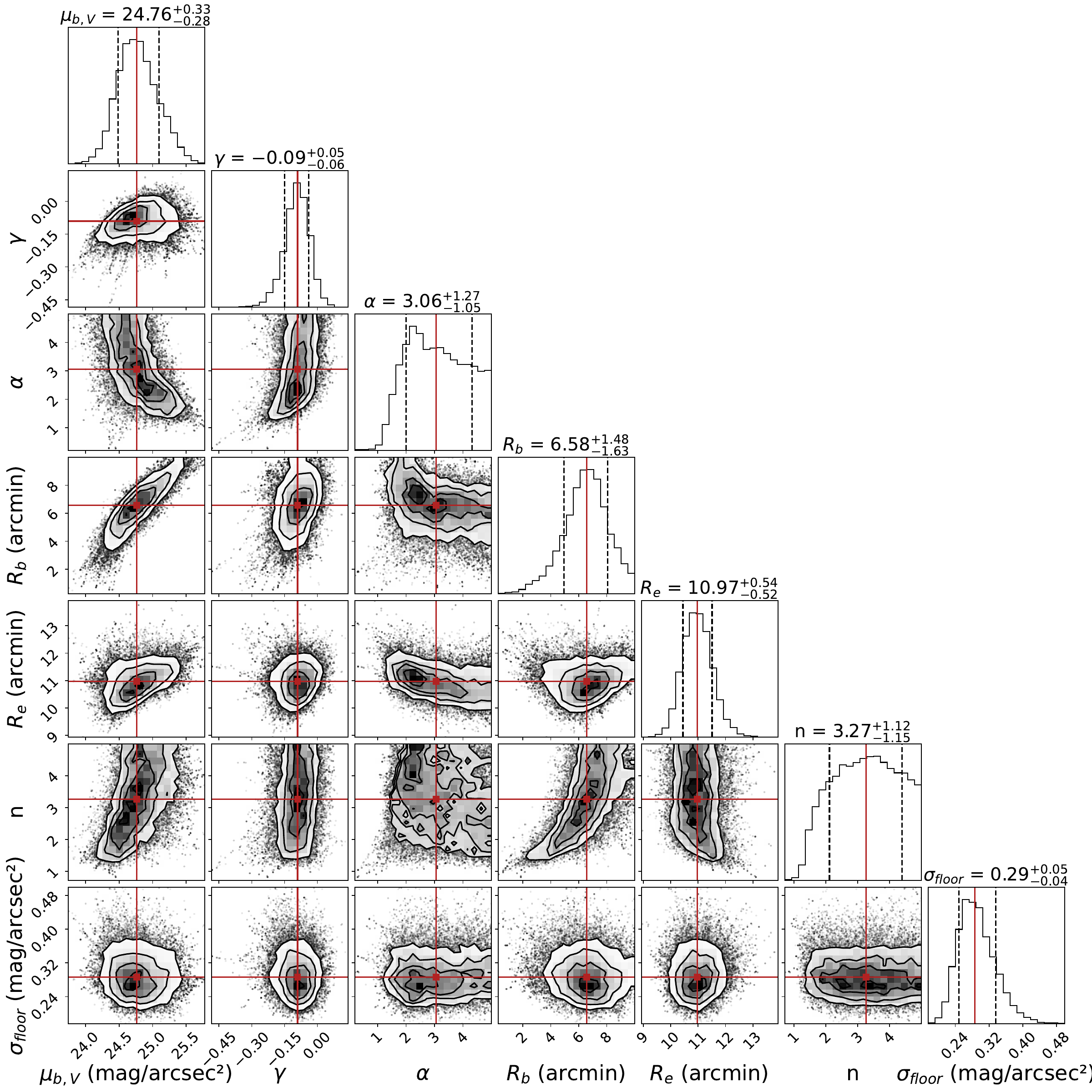}
    \end{subfigure}
    
    \vspace{0.3cm}
    
    \begin{subfigure}{\linewidth}
        \centering
        \includegraphics[width=0.5\linewidth]{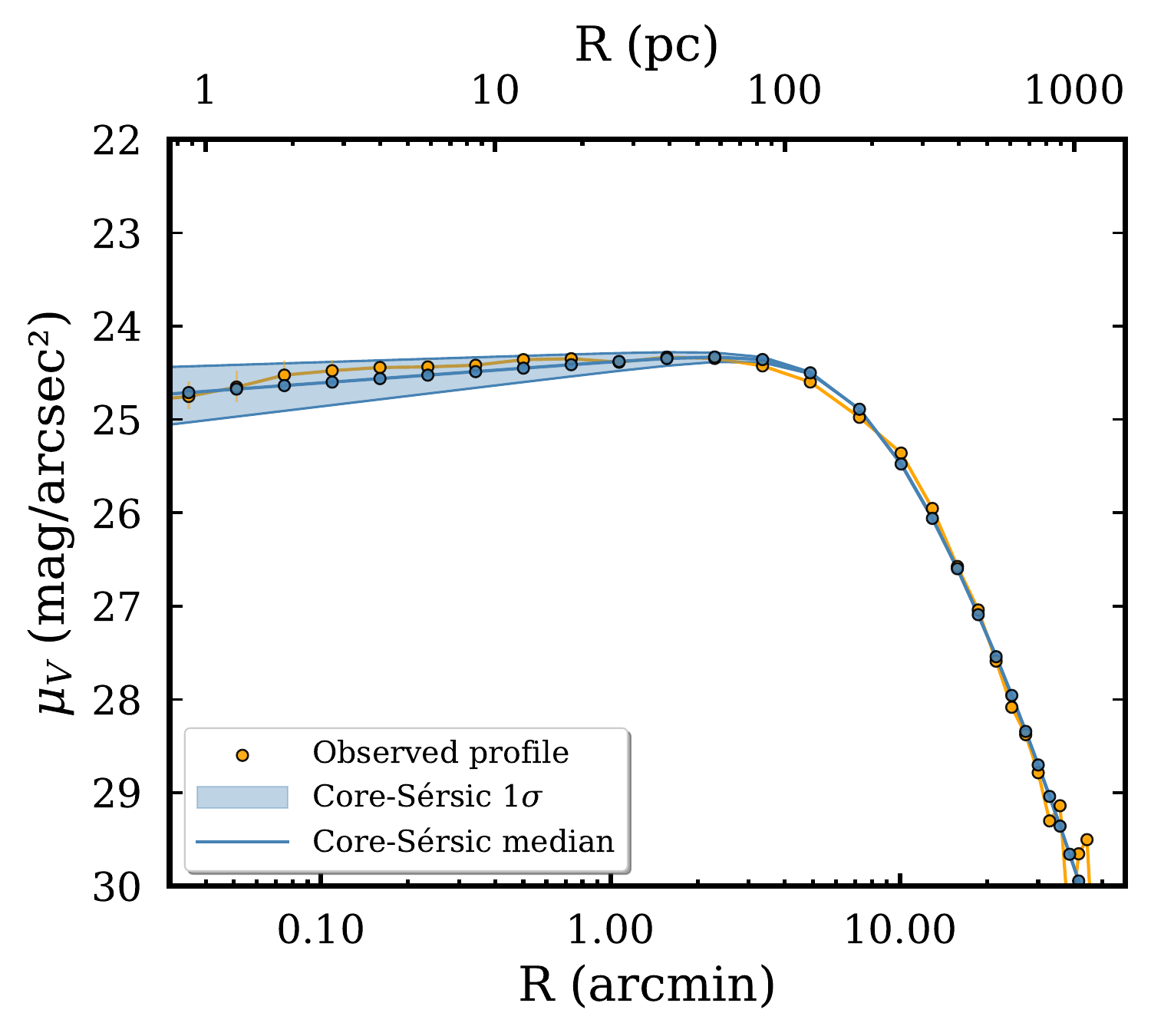}
    \end{subfigure}
    \caption{Core-S\'ersic modelling of the median surface brightness profile of Sculptor. Top: posterior distributions of the Core-S\'ersic parameters. Bottom: comparison between the observed profile and the best-fit model obtained. The observed profile is shown in yellow, and the best-fit model inferred from the MCMC is shown in blue.}
    \label{fig:Sculptor_full}
\end{figure*}

\begin{figure*}[h]
    \centering
    \begin{subfigure}{\linewidth}
        \centering
        \includegraphics[width=0.8\linewidth]{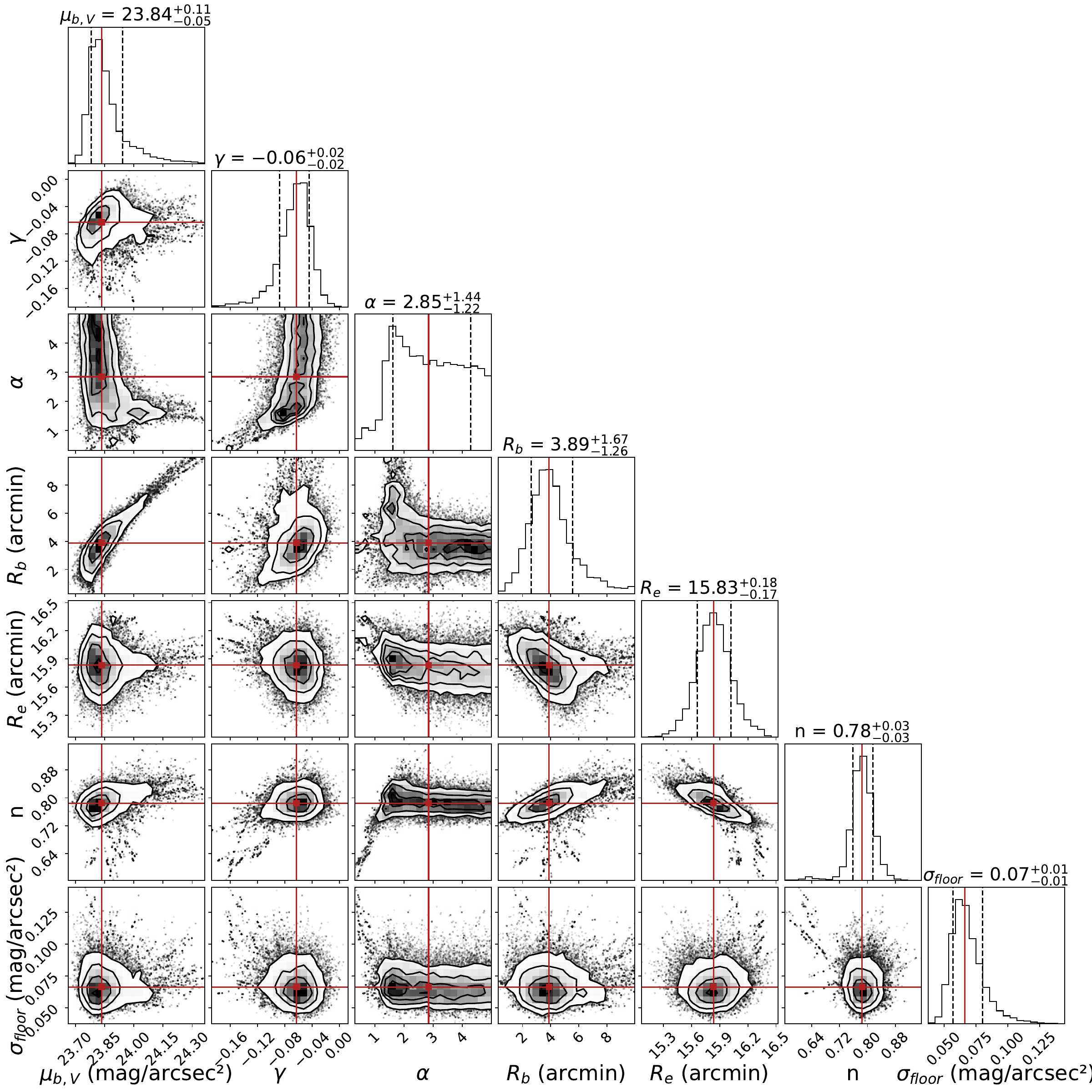}
    \end{subfigure}
    
    \vspace{0.3cm}
    
    \begin{subfigure}{\linewidth}
        \centering
        \includegraphics[width=0.5\linewidth]{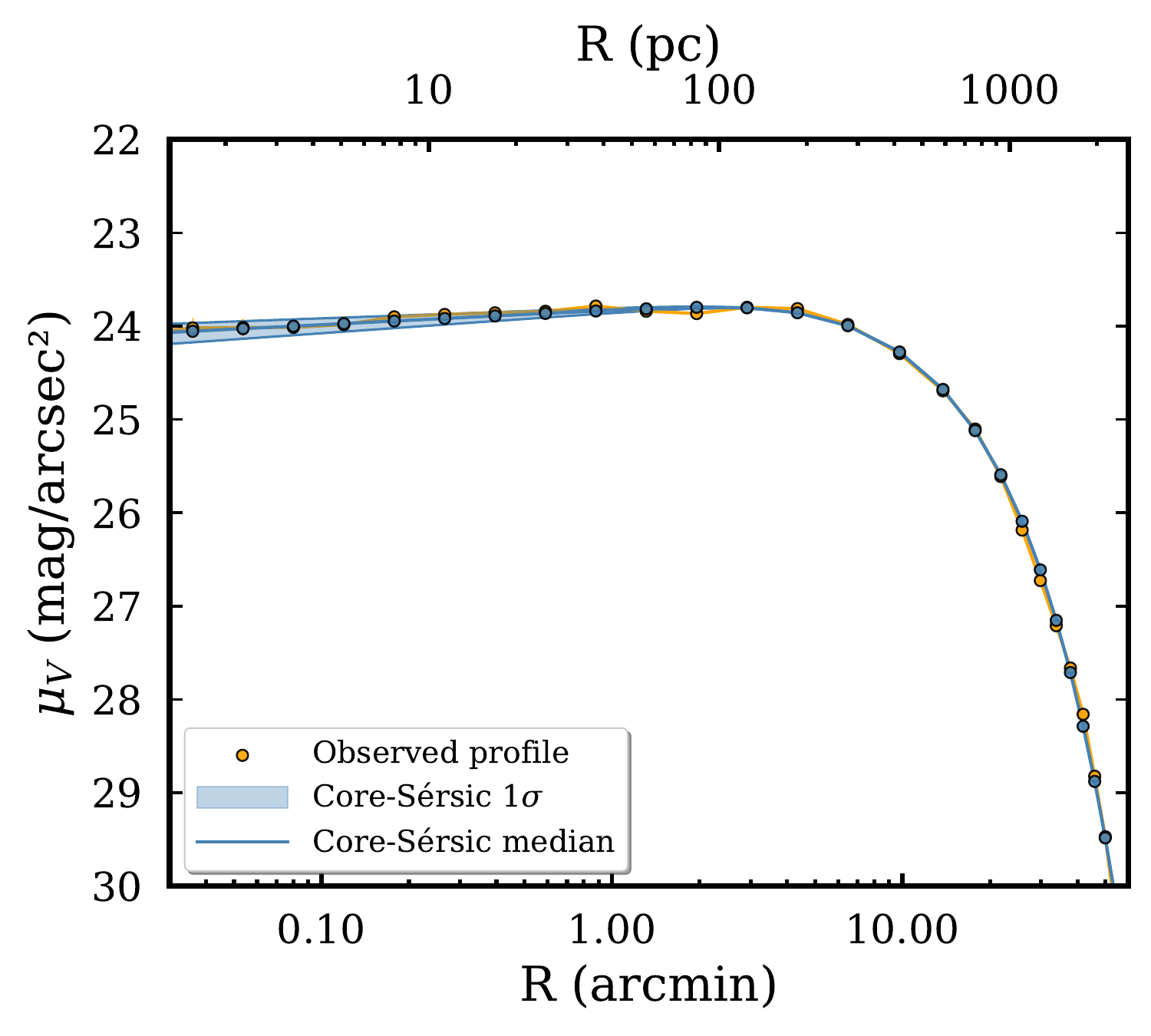}
    \end{subfigure}

    \caption{Core-S\'ersic modelling of the median surface brightness profile of Fornax. Top: posterior distributions of the Core-S\'ersic parameters. Bottom: comparison between the observed profile and the best-fit model obtained. The observed profile is shown in yellow, and the best-fit model inferred from the MCMC is shown in blue.}
    \label{fig:Fornax_full}
\end{figure*}

\end{appendix}
\end{document}